\documentclass[a4paper,runningheads]{llncs}
\pdfoutput=1
\usepackage[utf8]{inputenc}
\usepackage[T1]{fontenc}
\usepackage[dvipdfm]{graphicx}\usepackage{bmpsize}
\usepackage[hidelinks]{hyperref}
\usepackage{color}

\usepackage{orcidlink}

\usepackage[overlay]{textpos}
\usepackage{llncsstuffthm}

\usepackage[title]{appendix}
\usepackage{xspace}
\usepackage{array}
\usepackage{booktabs}
\usepackage{multirow}
\usepackage{fancyvrb}
\usepackage{fvextra}
\usepackage{amssymb}
\usepackage{amsfonts}
\usepackage{amsmath}
\usepackage{bm}
\mathchardef\hyph="2D
\usepackage{enumitem}

\newcommand{\oimp}{\Rightarrow}
\newcommand{\oimps}{{\Rightarrow}}
\newcommand{\graphedge}{\rightarrow}
\newcommand{\revimp}{\leftarrow}

\newcommand{\la}{\langle}
\newcommand{\ra}{\rangle}
\newcommand{\eqdef}{\;
\raisebox{-0.1ex}[0mm]{$ \stackrel{\raisebox{-0.2ex}{\tiny
      \textnormal{def}}}{=} $}\; }
\newcommand{\defname}[1]{\emph{#1}}
\newcommand{\name}[1]{\emph{#1}}
\newcommand{\f}[1]{\mathsf{#1}}

\newcommand{\D}{\f{D}}
\newcommand{\G}{\f{G}}
\newcommand{\fa}{\f{a}}
\newcommand{\fb}{\f{b}}

\newcommand{\fe}{\f{e}}
\newcommand{\ff}{\f{f}}
\newcommand{\fg}{\f{g}}
\newcommand{\fh}{\f{h}}

\newcommand{\fp}{\f{p}}

\newcommand{\SWIPL}{\textit{SWI-Prolog}\xspace}

\newcommand{\PIE}{\textit{PIE}\xspace}

\newcommand{\ProverN}{\textit{Prover9}\xspace}

\newcommand{\Metamath}{\textit{Metamath}\xspace}
\newcommand{\MM}{\name{Metamath}\xspace}

\newcommand{\CDTools}{\textit{CD Tools}\xspace}

\newcommand{\CDDC}{\name{CDDC}\xspace}

\newcommand{\Lukasiewicz}{{\L}ukasiewicz\xspace}

\newcolumntype{L}[1]{>{\raggedright\let\newline\\\arraybackslash\hspace{0pt}}p{#1}}
\newcolumntype{C}[1]{>{\centering\let\newline\\\arraybackslash\hspace{0pt}}p{#1}}

\newcommand{\txtcomment}[2]%
{{\color{blue}{\textbf{#1:} #2}}}

\newcommand\instanceOf{\mathrel{\ooalign{$\geq$\cr
      \hidewidth\raise.225ex\hbox{$\cdot\mkern7.0mu$}\cr}}}
\newcommand\strictInstanceOf{\mathrel{\ooalign{$>$\cr
      \hidewidth\raise.0ex\hbox{$\cdot\mkern7.0mu$}\cr}}}
\newcommand\subsumes{\mathrel{\ooalign{$\leq$\cr
      \hidewidth\raise.225ex\hbox{$\cdot\mkern2.0mu$}\cr}}}
\newcommand\variant{\mathrel{\ooalign{$=$\cr
      \hidewidth\raise.7ex\hbox{$\cdot\mkern4.5mu$}\cr}}}

\newcommand{\tup}[1]{\boldsymbol{#1}}
\newcommand{\Rs}{\tup{R}}
\newcommand{\Us}{\tup{U}}
\newcommand{\Vs}{\tup{V}}

\newcommand{\mmname}[1]{\texttt{#1}}

\newcommand{\pro}{\rightarrow}

\newcommand{\setmm}{\textit{set.mm}\xspace}
\newcommand{\mmexe}{\textit{metamath.exe}\xspace}

\newcommand{\val}{\f{val}}
\newcommand{\refcount}{\f{ref}}
\newcommand{\savval}{\f{sav}}
\newcommand{\base}{\mathcal{B}}
\newcommand{\clauses}{\mathcal{F}}

\newcommand{\mgt}{\f{mgt}}
\newcommand{\grammarmgt}{\f{grammar\hyph mgt}}
\newcommand{\shallowmgt}{\f{shallow\hyph mgt}}

\newcommand{\rname}[1]{\textsc{#1}}

\newcommand{\mgu}{\f{mgu}}

\newcommand{\AP}{\rname{AppPar}}

\newcommand{\Start}{\f{Start}}

\newcommand{\arity}{\f{arity}}

\newcommand{\pdnet}{\f{PDNet}}

\newcommand{\PDNET}{PDNet\xspace}
\newcommand{\PDNETs}{PDNets\xspace}

\newcommand{\ATT}{}

\newcommand{\xmin}{$\land$}
\newcommand{\xmax}{$\lor$}
\newcommand{\xavg}{$\overline{x}$}
\newcommand{\xmed}{$\tilde{x}$}

\newcommand{\nonlin}{\f{nl}}
\newcommand{\varoccs}{\f{voccs}}
\newcommand{\argmult}{\f{vmult}}

\newcommand{\N}{\f{N}}

\newcommand{\height}{\f{h}}

\newcommand{\UpSigma}{\mathrm{\Sigma}}

\newcommand{\SETCORE}{\textsc{SetCore}\xspace}

\newcommand{\MiniSet}{\textsc{MiniSet}\xspace}
\newcommand{\MiniDag}{\textsc{MiniDag}\xspace}
\newcommand{\MiniTrp}{\textsc{MiniTrp}\xspace}
\newcommand{\MiniXC}{\textsc{MiniGrc}\xspace}

\newcommand{\TRP}{\f{TRP}}

\newcommand{\hcmed}{\multicolumn{1}{c}{\xmed}}
\newcommand{\hcmin}{\multicolumn{1}{c}{\xmin}}
\newcommand{\hcmax}{\multicolumn{1}{c}{\xmax}}
\newcommand{\hcavg}{\multicolumn{1}{c}{\xavg}}
\newcommand{\hc}[1]{\multicolumn{1}{c}{#1}}

\newcommand{\xset}{\textsc{Set}}
\newcommand{\xmm}{\textsc{MS}}
\newcommand{\xsc}[1]{\scalebox{0.7}{#1}}

\newcommand{\xitem}[1]{\item[#1]}

\newenvironment{coldesc}[1]
               {\begin{list}{}{
                     \setlength{\labelwidth}{#1}
                     \setlength{\leftmargin}{#1}
                     \setlength{\itemindent}{0.0cm}                     
               }}
               {\end{list}}

\newcounter{obscounter}

\newenvironment{obslist}
               {\begin{enumerate}[label={$\blacktriangleright$
                       \textbf{O\arabic*.}},
                     ref=\textbf{O\arabic*},
                     align=left,
                     itemindent=17pt,
                     leftmargin=0pt
                   ]
                   \setcounter{enumi}{\value{obscounter}}}
               {\setcounter{obscounter}{\value{enumi}}\end{enumerate}}

\makeatletter
\renewcommand\section{\@startsection{section}{1}{\z@}%
                       {-18\p@ \@plus -4\p@ \@minus -4\p@}%
                       {12\p@ \@plus 4\p@ \@minus 4\p@}%
                       {\normalfont\large\bfseries\boldmath
                        \rightskip=\z@ \@plus 8em\pretolerance=10000 }}
\renewcommand\subsection{\@startsection{subsection}{2}{\z@}%
                       {-18\p@ \@plus -4\p@ \@minus -4\p@}%
                       {8\p@ \@plus 4\p@ \@minus 4\p@}%
                       {\normalfont\normalsize\bfseries\boldmath
                        \rightskip=\z@ \@plus 8em\pretolerance=10000 }}
\renewcommand\subsubsection{\@startsection{subsubsection}{3}{\z@}%
                       {-18\p@ \@plus -4\p@ \@minus -4\p@}%
                       {-0.5em \@plus -0.22em \@minus -0.1em}%
                       {\normalfont\normalsize\bfseries\boldmath}}
\renewcommand\paragraph{\@startsection{paragraph}{4}{\z@}%
                       {-12\p@ \@plus -4\p@ \@minus -4\p@}%
                       {-0.5em \@plus -0.22em \@minus -0.1em}%
                       {\normalfont\normalsize\itshape}}
\renewcommand\paragraph{\@startsection{paragraph}{4}{\z@}%
                       {-4\p@ \@plus -4\p@ \@minus -4\p@}%
                       {-0.5em \@plus -0.22em \@minus -0.1em}%
                       {\normalfont\normalsize\itshape}}
\renewcommand\subsubsection{\@startsection{subsubsection}{3}{\z@}%
                       {-6\p@ \@plus -2\p@ \@minus -4\p@}%
                       {-0.5em \@plus -0.22em \@minus -0.1em}%
                       {\normalfont\normalsize\bfseries\boldmath}}
\renewcommand\section{\@startsection{section}{1}{\z@}%
                       {-8\p@ \@plus -4\p@ \@minus -4\p@}%
                       {6\p@ \@plus 4\p@ \@minus 4\p@}%
                       {\normalfont\large\bfseries\boldmath
                        \rightskip=\z@ \@plus 8em\pretolerance=10000 }}

\def\@listI{\leftmargin\leftmargini
            \parsep 0\p@ \@plus1\p@ \@minus\p@
            \topsep 2\p@ \@plus4\p@ \@minus\p@
            \itemsep0\p@}
\let\@listi\@listI
\@listi
\makeatother

\begin{document}

\addtolength{\textheight}{0.5cm}

\setlength{\abovedisplayskip}{0.6ex plus0.2ex}%
\setlength{\belowdisplayskip}{0.6ex plus0.2ex}%
\setlength{\abovedisplayshortskip}{0pt}%
\setlength{\belowdisplayshortskip}{0pt}%

\setlength{\textfloatsep}{10pt}

\title{Mathematical Knowledge Bases as Grammar-Compressed Proof Terms:\\
  Exploring Metamath Proof Structures}

\author{Christoph Wernhard~\inst{1} \and Zsolt
  Zombori~\inst{2,3}}

\authorrunning{C.~Wernhard \and Z. Zombori}
\titlerunning{Mathematical Knowledge Bases as Grammar-Compressed Proof Terms}

\institute{University of Potsdam, Germany%
  \and HUN-REN Alfr\'ed R\'enyi Institute of Mathematics,
  Hungary%
  \and E{\"o}tv{\"o}s Lor\'{a}nd University,
  Budapest, Hungary}

\maketitle

\begin{abstract}
Viewing formal mathematical proofs as logical terms provides a powerful and
elegant basis for analyzing how human experts tend to structure proofs and how
proofs can be structured by automated methods. We pursue this approach by (1)
combining proof structuring and grammar-based tree compression, where we show
how they are inherently related, and (2) exploring ways to combine human and
automated proof structuring. Our source of human-structured proofs is
Metamath, which, based on condensed detachment, naturally provides a view of
proofs as terms. A knowledge base is then just a grammar that compresses a set
of gigantic proof trees. We present a formal account of this view, an
implemented practical toolkit as well as experimental results.
\end{abstract}

\begin{textblock}{12.2}(0,9.5)
  \scriptsize
  \noindent \rule{5em}{0.5pt}\\[1pt]
  Funded by the Deutsche Forschungsgemeinschaft (DFG, German
  Research Foundation) -- Project-ID~457292495. The authors would like to
  thank the Federal Ministry of Education and Research and the state
  governments (\url{https://www.nhr-verein.de/unsere-partner}) for supporting
  this work as part of the joint funding of National High Performance
  Computing (NHR). Funded by the Hungarian Artificial Intelligence National
  Laboratory Program (RRF-2.3.1-21-2022-00004) as well as the ELTE TKP
  2021-NKTA-62 funding scheme. This article is based upon
  work from the action CA20111 EuroProofNet supported by COST (European
  Cooperation in Science and Technology).
\end{textblock}

\section{Introduction}
\label{sec-introduction}

The very essence of formalized mathematics may be characterized as starting
from a few axioms and using a few inference rules to derive theorems. A
verifier can then check a given derivation and a theorem prover can search for
a derivation of a given conjectured theorem. However, derivations can be very large
and also, even for limited size, the number of possible derivations is very
large. Thus, ``something else'' is assumed that allows to distinguish theorems
of interest and to guide proof search.
Forms of this ``something else'' are usually sought outside of the axiomatic
approach.
In automated proving, for example, with heuristics, notions of redundancy and
saturation, or coupling with machine learning. In interactive proving, for
example, by assigning names to distinguished formulas.

The particular form of this ``something else'' considered here is
\emph{compression of proof structures}, which is already inherent in certain
variants of the axiomatic approach such as \MM
\cite{megill:1995,metamath:book,metamath:website}, a successful computer
language for archiving, verifying, and studying mathematical proofs. \MM
continues a thread from \Lukasiewicz and Tarski who investigate
axiomatizations of propositional logics within a first-order meta-level
framework \cite{luk:tarski:aussagenkalkuel:1930}, via Meredith, who refines
this with his \name{condensed detachment} by the implicit use of most general
unifiers instead of explicit substitutions and a view on proof structures as
terms with a DAG representation
\cite{prior:logicians:1956,prior:formal:logic:1962,hindley:meredith:cd:1990,mccune:wos:cd:1992,ulrich:legacy:2001,cwwb:article:2024}.

\MM proofs are in essence grammar-compressions
\cite{lohrey:treerepair:2013,lohrey:survey:2015} of proof terms built
from just two primitive inference rules: condensed detachment (modus ponens
with unification) and condensed generalization (quantifier introduction).
A non-cyclic tree grammar with a single production for each nonterminal
provides a compressed representation of a set of proof terms.
Repeated \emph{patterns}, such as the two occurrences of $\fg(\fh(\_))$ in
$\ff(\fg(\fh(\fa)),\fg(\fh(\fb)))$ can be factored by non-terminals with
parameters, such as in the grammar $\{\fp(V) \pro \fg(\fh(V)),\;
\Start \pro \ff(\fp(\fa),\fp(\fb))\}$. The case with no parameter represents the
factoring of a subtree as in a DAG.

With this form of compression of proof terms we can model the way in which
mathematical knowledge is structured in \MM. A grammar production corresponds
to the proof of a theorem. The theorem formula is, from the first-order
perspective (the ``meta-level'' at which mathematics is expressed in \MM) a
definite clause. The length of its body is the number of parameters of the
corresponding production. The production's nonterminal serves as theorem name
and as a function symbol in other proof terms.
Given a grammar, axiom clauses and a proof term built from nonterminals, axiom
symbols and variables, a most general clause proven by the proof term can be
uniquely determined.

The largest \MM database is the \name{Metamath Proof Explorer}, also called
\setmm, with about 44,000 theorems on various mathematical topics, developed
with Zermelo-Fraenkel set theory. Such a knowledge base now appears as a
single tree grammar with a production for each theorem. The expanded value of
a nonterminal is a large tree whose inner nodes represent applications of the
two primitive inference rules, and whose leaves represent instances of axioms.
Can the compression view help us to understand how human experts structure
mathematical knowledge? Are such principles useful for guiding proof search?
Can we apply automated techniques such as tree compression algorithms, e.g.,
\cite{lohrey:treerepair:2013}, to improve on the human-made structuring of
\setmm? Such questions lead us to investigate combinations of proof
structuring and grammar-based tree compression, and to explore combinations of
human and automated proof structuring.

\paragraph{Structure of the Paper.}
We formally generalize the framework of condensed detachment by allowing
parameters in
proof terms (Sect.~\ref{sec-formal}) and show how this captures the
correspondence between tree grammars and proofs as terms, adequately for \MM
(Sect.~\ref{sec-grammar-compressed-trees}). We then take an empirical look at
various properties of a large fragment of \setmm from the grammar perspective
(Sect.~\ref{sec-properties-kbs}). After describing implemented proof tree
compression techniques, we compare their results for a small subset of \setmm
with the original structuring in \setmm (Sect.~\ref{sec-comparing-mini}). By
combining human structuring and machine compression we strive for new lemmas
that are useful with large subsets of \setmm (Sect.~\ref{sec-humancompress}).
Finally, we discuss potential applications and related works, pointing out
distinguishing features of our approach and
conclude with an outlook (Sect.~\ref{sec-conclusion}).

\paragraph{Implementation.}
We use our system \CDTools \cite{cw:ccs:2022,cw:sgcd:2024,rwzb:lemmas:2023},
written in \SWIPL \cite{swiprolog} and embedded in \PIE
\cite{cw:pie:2016,cw:pie:2020}, which addresses experimenting with condensed
detachment. It now comprises an advanced \MM interface and tree compression
methods. The system and code for additional experiments are available from
\url{http://cs.christophwernhard.com/cdtools/}. Experiments with \setmm
refer to \url{https://github.com/metamath/set.mm}, commit 8cf01a7, 10 Jan,~2025.

\section{Condensed Detachment and Definite Clauses}
\label{sec-formal}

\begin{table}[t]
  \caption{The rules of the \CDDC inference system.}
  \vspace{-2pt}
  \label{tab-cddc}
  \centering
  $\begin{array}{lc}
    \multirow{2}{*}{\raisebox{-3pt}{\rname{App}\;\;}} &
    p :: A \revimp B_1 \land \ldots \land B_n \hspace{1em} d_1 : B_1' \revimp \Rs_1
    \;\; \ldots\;\; d_n : B_n' \revimp \Rs_n
    \\\cmidrule{2-2}
    & p(d_1, \ldots, d_n) : (A \revimp \Us)\sigma,\\[2pt]
    & \text{where } \sigma = \mgu(\{\{B_1,B_1'\}, \ldots ,\{B_n,B_n'\},
    \{\Us,\Rs_1,\ldots,\Rs_n\}\})\\
    & \multicolumn{1}{l}{\text{and the first premise is in } \base}
  \end{array}$

  \smallskip
  
  $\begin{array}[t]{lc}
 \multirow{2}{*}{\raisebox{-3pt}{\rname{Par}\;\;}}\\\cmidrule{2-2}
 & V_i : u_i \revimp \Us
  \end{array}$
  \hspace{3.9cm}
  $\begin{array}[t]{lc}
    \multirow{2}{*}{\raisebox{-3pt}{\rname{Ins}\;\;}} &
    d : A \revimp \Rs\\\cmidrule{2-2}
    & d : (A \revimp \Rs)\sigma\\[2pt]
  \end{array}$
\end{table}

Central to our approach is a view of proofs as terms, made precise as follows.

\vspace{-4pt}
\begin{defn}[Proof Term]
  A \defname{proof term} $d$ is a term built from function symbols~$p$ with
  arity $\arity(p) \geq 0$, called \defname{presupposition names}, and
  \defname{parameters}\linebreak $V_1, V_2, \ldots$. A proof term without parameters is
  called \defname{ground}. A proof term without multiple occurrences of the
  same parameter is called \defname{linear}. We write a proof term $d$ as
  $d[V_1,\ldots,V_k]$ to indicate that parameters $V_1,\ldots,V_k$ occur in
  $d$. For proof terms $d_1,\ldots,d_k$, then $d[d_1,\ldots,d_k]$ denotes $d$
  with the $V_i$ replaced by~$d_i$.
\end{defn}
\vspace{-4pt}

\noindent
The \name{presupposition names} in a proof term label formulas that play the
\emph{the role of axioms}, which can be actual axioms or lemmas that have been
proven before.
A proof term is related to a \name{most general theorem (MGT)}, a unique most
general formula that is proven as described by the proof term from given
presuppositions. This view originates in the framework of \name{condensed
  detachment}, which we present in a generalized version by means of an
inference system \name{CDDC} (\name{Condensed Detachment for Definite
  Clauses}). \CDDC defines the \name{proves} relation $d : F$ between
a proof term~$d$ and a formula~$F$. The formula is a definite first-order
clause (briefly called \name{clause} here) $A \revimp B_1 \land \ldots \land B_n$, $n
\geq 0$, where all atoms have the same unary predicate ``provable'', such that
we actually represent these atoms just by their argument term. \Metamath
axioms and theorems can be understood as such first-order clauses, where
logical connectives and predicates at the object-level are represented by
first-order function symbols. Our inference system \CDDC operates on such
pairs $d : F$, where the proof structure is \emph{reified as a term} $d$
instead of just implicit in the derivation structure.
The role of the inference system is to provide a foundation, a specification
of constraints. It is not intended as a recipe for proof search or structuring
proofs. A natural approach to search is by enumerating proof terms in
combination with unification \cite{cw:sgcd:2024,cw:ccs:2022,rwzb:lemmas:2023},
rooted in
\cite{prawitz:1960:improved,prawitz:1969:advances,loveland:1968,pttp,bibel:atp:1987,bibel:otten:2020}.
Table~\ref{tab-cddc} shows the three rules of \defname{CDDC}:
\name{presupposition application} (\rname{App}), \name{parameter recording}
(\rname{Par}) and \name{instantiation} (\rname{Ins}). It operates on
two kinds of statements:
  \begin{itemize}
  \item \defname{Presupposition-statements} $p :: F$, where $p$ is
    a presupposition name and $F$ is a clause with body length $\arity(p)$.
  \item \defname{Proves-statements} $d : F$, where $d$ is a proof term and $F$
    is a clause whose body length is at least the maximum index $i$ of parameters
    $V_i$ occurring in~$d$.
  \end{itemize}

\enlargethispage{4pt}
  
\noindent  
\CDDC assumes a given \defname{presupposition base} (briefly
\name{base})~$\base$, that is, a set of presupposition-statements with
pairwise distinct presupposition names. A base is said to be \defname{for} a
proof term~$d$ if it contains a presupposition-statement for all
presupposition names occurring in~$d$. We further assume that the parameters
that may occur in a proof term are $V_1, \ldots, V_k$ and associate dedicated
formula variables $u_1, \ldots, u_k$ with these, writing $\Us$ as shorthand
for the body $u_1 \land \ldots \land u_k$. In rule \rname{App} it is assumed
that the premises have disjoint sets of variables, which are also disjoint
from those of $\Us$. This can be achieved by renaming of variables, if
necessary.
We write $\mgu(S)$ for the \name{most general unifier} of $S = \{T_1, \ldots,
T_n\}$, a set $S$ of sets $T_i$ of terms. It is the most general substitution
$\sigma$ such that for all $i \in \{1,\ldots,n\}$ with $T_i$ not empty
$T_i\sigma$ is a singleton. If such a substitution exists, then $\mgu(S)$ is
\defname{defined}. We assume w.l.o.g. that for $\mgu(S)$ its domain and the
set of variables occurring in its range are disjoint subsets of the set of
variables in the unified terms \cite[Rem.~4.2]{eder:subst:1985}.
We define our notion of \name{MGT} as follows.
\begin{defn}[Most General Theorem -- MGT]
\label{def-mgt}  
An $\AP$-deduction of a proves-statement $S = (d : F)$ from a presupposition
base $\base$ is a sequence $S_1,\ldots,S_k$ of presupposition- and
proves-statements such that $S_k = S$ and each $S_i$ is either a member of
$\base$ or a proves-statement obtained from premises preceding $S_i$ with
rule $\rname{App}$ or $\rname{Par}$.
If there is an $\AP$-deduction of a statement $d[V_1,\ldots,V_k] : A
\revimp B_1 \land \ldots \land B_k$, we say that
$\mgt_\base(d[V_1,\ldots,V_k])$
is \defname{defined} and
\[\mgt_\base(d[V_1,\ldots,V_k])\; =\; A \revimp B_1 \land \ldots \land B_k.\]
\end{defn}

Since, for given $\base, d$, the clause $\mgt_\base(d)$ is unique up to
renaming of variables, we call that clause informally \name{the MGT} of $d$.
Note that the MGT of a proof term with parameters has for each parameter $V_i$
a corresponding body literal $B_i$. The MGT of a ground proof term is thus a
positive unit clause.

\CDDC is best illustrated with some special cases. If the presupposition $p$
has an empty body, rule \rname{App}, \name{presupposition application},
specializes as follows.
{\par\centering\small
  \smallskip
$\begin{array}{c}
  p :: A\\\midrule
  p : A \revimp \Us
  \end{array}$\hspace{8pt}\rname{App}\par
  \smallskip  
}
\noindent
For example, if $\base$ includes $\mmname{ax-1} :: (x \oimp (y \oimp x))$, we
can infer the proves-statement $\mmname{ax-1} : (x \oimp (y \oimp x)) \revimp
\Us$, where $\oimp$ is a binary function symbol that represents implication at
the object-level. The added body $\Us$ is the conjunction $u_1 \land \ldots
\land u_k$ of one formula variable $u_i$ for each proof term parameter $V_i$
that may occur in a larger considered proof term. Rule \rname{Par},
\name{parameter recording}, effects that for all occurrences of a parameter
$V_i$ in the proof term the head of the clause that is ``proven'' by the
parameter is identified with the corresponding variable $u_i$ in $\Us$. In \MM
the stated user-specified theorem formulas can be \emph{strict} instances of
the MGTs of their proofs. Rule \rname{Ins}, \name{instantiation}, where
$\sigma$ is an arbitrary substitution that respects certain restrictions to
ensure that the formula in the conclusion is still well-formed
\cite{megill:1995,metamath:book}, has the purpose to model this. Continuing
our example, with an \rname{Ins} step we can infer, e.g., $\mmname{ax-1} : ((x
\oimp y) \oimp (z \oimp (x \oimp y))) \revimp \Us$.

\enlargethispage{12pt}

We now consider a particularly important specialization of \rname{App} that
represents \name{condensed detachment}, one of two primitive proof
constructors of \setmm.
{\par\centering\small
  \smallskip
  $\begin{array}{c}
  \f{D} :: y \revimp (x \oimp y) \land x\hspace{1em}
  d_1 : B_1' \revimp \Rs_1\hspace{1em} d_2 : B_2' \revimp \Rs_2\\\midrule
  \D(d_1, d_2) : (y \revimp \Us)\sigma,
  \end{array}$\hspace{8pt}\rname{App}\par\smallskip
}
\noindent
where $\sigma = \mgu(\{\{(x \oimp y), B_1'\}, \{x,B_2'\},
\{\Us,\Rs_1,\Rs_2\}\}).$ We assume that $\base$ contains the
presupposition-statement $\f{D} :: y \revimp (x \oimp y) \land x$. Proof terms
$d_1, d_2$ prove the major and the minor premise of the detachment. In \setmm, $\D$ is
axiom \mmname{ax-mp}, with reversed order of parameters. Unification of
$\{\Us,\Rs_1,\Rs_2\}$ effects that constraints on the $u_i$ variables (which
correspond to parameters $V_i$) imposed with the subproofs $d_1, d_2$ get
propagated to the body of the rule's conclusion.
The second primitive proof constructor of \setmm is $\G$, \name{condensed
  generalization}, represented by the presupposition-statement $\G ::
\forall(x, y) \revimp y$, where $\forall$ is a binary function symbol that
represents universal quantification at the object-level. In \setmm, $\G$ is
axiom \mmname{ax-gen}. Here are some example MGTs.

\begin{examp}
  \label{examp-mgt}
  Let $\base = \{\f{D} :: y \revimp (x \oimp y) \land x,\; \mmname{ax-1} ::
  x_1\oimp (x_2\oimp x_1),\; \mmname{ax-2} :: ((x_1\oimp (x_2\oimp x_3))\oimp
  ((x_1\oimp x_2)\oimp (x_1\oimp x_3)))\}$. \mmname{ax-1} is known as
  \name{Simp} and as \name{K}, \mmname{ax-2} as \name{Frege} and as \name{S}
  \cite{ulrich:legacy:2001}. Both are in \setmm. Then
  \begin{enumerate}[left= 0pt,label=(\roman*)]
    \item $\mgt_\base(\mmname{ax-1}) = (x_1\oimp (x_2 \oimp x_1))$.
    \item $\mgt_\base(\D(\mmname{ax-1},\mmname{ax-1})) = (x_1 \oimp (x_2 \oimp (x_3 \oimp  x_2)))$.
    \item $\mgt_\base(\D(V_1,\mmname{ax-1})) = x_1 \revimp ((x_2\oimp (x_3\oimp x_2))\oimp x_1)$.
    \item $\mgt_\base(\D(\mmname{ax-1}, \D(V_1, \mmname{ax-1}))) = (x_1\oimp x_2) \revimp ((x_3\oimp
      (x_4\oimp x_3))\oimp x_2)$.
  \item  $\mgt_\base(\D(\D(\mmname{ax-2}, \mmname{ax-2}), \mmname{ax-2}))$ is undefined.      
  \item $\mgt_\base(\D(\D(\mmname{ax-2}, V_1), \mmname{ax-2})) = ((x_1\oimp (x_2\oimp x_3))\oimp x_4) \revimp$\\\hspace*{\fill} $((x_1\oimp (x_2\oimp x_3))\oimp (((x_1\oimp x_2)\oimp (x_1\oimp x_3))\oimp x_4))$.
  \item $\mgt_\base(\D(\D(\mmname{ax-2},\mmname{ax-1}),\mmname{ax-2})) = ((x_1\oimps (x_2\oimps x_3))\oimp (x_1\oimps (x_2\oimps x_3)))$.
  \end{enumerate}
\end{examp}

There is a subtlety concerning \emph{nonlinear} proof terms. Variables $u_i$
(which correspond to a parameter $V_i$) are constrained with respect to
\emph{each occurrence} of $V_i$ in the proof term. Unification of
$\{\Us,\Rs_1,\ldots,\Rs_n\}$ in rule \rname{App} effects that these
constraints are merged, such that the instance of $u_i$ in the final MGT
simultaneously fulfills the constraints imposed on each occurrence of $V_i$.
This may lead to a mismatch between two ways of determining a MGT: (1) From a
proof term with parameters and the MGTs of proof terms intended to substitute
these; (2) From the proof term after performing the substitution. There are
cases where (1) yields a strict instance of the latter, or even where for (1)
the MGT is undefined although for (2) it is defined. This is not just a
marginal phenomenon: The proof terms for about one third of the theorems in
\setmm are nonlinear. The
following proposition states coincidence of (1) and (2) for linear proof
terms, and the instance relationship for the general case. We write $F \instanceOf
F'$ for $F$ \name{is an instance of} $F'$.
\begin{prop}
  \label{prop-inst}
  Let $d[V_1,\ldots,V_k]$ be a proof term, let $\base$ be a presupposition
  base for $d$ and let $A, B_1, \ldots, B_k$ be atoms such that
  $\mgt_{\base}(d[V_1,\ldots,V_k]) = A \revimp B_1 \land \ldots \land B_k$.
  Let $d_1, \ldots, d_k$ be ground proof terms such that $\base$ is also a
  base for them, let $B'_1, \ldots, B'_k$ be atoms that do not share variables
  among each other nor with $A,\linebreak B_1, \ldots, B_k$ such that
  $\mgt_{\base}(d_1) = B'_1, \ldots, \mgt_{\base}(d_k) = B'_k$ and such that
  $\sigma = \mgu(\{\{B_1,B'_1\}, \ldots, \{B_k,B'_k\}\})$ is defined. If $d$
  is linear, then $\mgt_{\base}(d[d_1,\ldots,d_k]) = A\sigma.$ In the general
  case, also for nonlinear $d$, the following weaker statement holds:
  $A\sigma \instanceOf \mgt_{\base}(d[d_1,\ldots,d_k])$.
\end{prop}

\begin{examp}
  \label{examp-inst-nonlinear}
  Let $\base$ be as in Examp.~\ref{examp-mgt}. Let $d[V_1] = \D(V_1,
  \D(\D(V_1, \mmname{ax-1}), \mmname{ax-1}))$ and let $d_1 = \mmname{ax-1}$.
  Proof term $d$ is nonlinear and Prop.~\ref{prop-inst} applies with the
  following values:
  $A = ((y_1 \oimp  (y_2 \oimp  y_1)) \oimp  (y_3 \oimp (y_4 \oimp  y_3)))$;
  $B_1 = ((y_3 \oimp (y_4 \oimp  y_3)) \oimp
  ((y_1 \oimp  (y_2 \oimp  y_1)) \oimp (y_3 \oimp  (y_4 \oimp y_3))))$;
  $B'_1 = (x_1\oimp (x_2\oimp x_1))$;
  $\sigma = \{x_1 \mapsto (y_3\oimps (y_4\oimps y_3)),\; x_2 \mapsto
  (y_1\oimps (y_2\oimps y_1))\}$.
  We have $\mgt_{\base}(d[d_1]) : A'$, where $A' = (z_1\oimp (z_2\oimp
  (z_3\oimp z_2)))$. Atom $A\sigma = A$ is a strict instance of $A'$: $A\sigma
  = A = A'\rho$, where $\rho = \{z_1 \mapsto (y_1\oimp (y_2\oimp y_1)),\; z_2
  \mapsto y_3,\; z_3 \mapsto y_4\}$.
\end{examp}

\section{Grammar-Compressed Trees and Knowledge Bases}
\label{sec-grammar-compressed-trees}

We apply general techniques of grammar-based tree compression
\cite{lohrey:treerepair:2013,lohrey:survey:2015} to proof terms, where a set
of terms is compactly described by a specific kind of grammar.\linebreak
A \defname{production} $P$ has the form $p(V_1,\ldots,V_n) \pro d$, where $p$
is a \defname{nonterminal} with arity $n \geq 0$, $d$ is a term (i.e., a proof
term), and $V_1,\ldots,V_n$ are \defname{parameters}, exactly those that occur
in $d$. $p(V_1,\ldots,V_n)$ and $d$ are called LHS and RHS of $P$.
A \defname{proof grammar}~$G$ is a finite sequence $\la P_1,\ldots,P_k \ra$ of
productions with pairwise different nonterminals $p_1,\ldots,p_k$, such that
if $p_i$ occurs in the RHS of $P_j$, then $i < j$. The term obtained by
exhaustively expanding a LHS $p(\Vs)$
with the productions of $G$ is denoted
by $\val_G(p(\Vs))$. It is a term with parameters $\Vs$.\footnote{Our
\name{proof grammars} generalize the grammars considered in
\cite{lohrey:treerepair:2013} in several respects: we permit nonlinear and
single-parameter RHSs, we do not require a distinguished \name{start}
nonterminal and we consider $\val$ for arbitrary LHSs.}
\name{Terminals} are just those RHS symbols that are in no LHS.
A production is \defname{linear} if its RHS has no multiple occurrences of the
same parameter. A grammar is \defname{linear} if all its productions are
linear.
The \defname{size}~$|d|$ of a term~$d$ is the number of its edges when viewed
as tree, e.g., $|V_1| = |\mmname{ax-1}| = 0$, $|\G(\mmname{ax-1})| = 1$,
$|\D(\G(\mmname{ax-1}),\mmname{ax-2})| = 3$. The \defname{size} of a
production is the size of its RHS. The size~$|G|$ of a grammar~$G$ is the sum
of the sizes of its productions.
The \defname{arity} of a grammar is the maximum arity of its nonterminals. A
grammar with arity $0$ is called a \name{DAG compression}. Any term has a
unique DAG compression with minimal size.

\vspace{-3pt}
\begin{examp}\label{examp-grammar}
  Let $d = \D(\mmname{ax-1}, \D(\mmname{ax-1}, \D(\D(\mmname{ax-1},
  \mmname{ax-1}), \D(\mmname{ax-1}, \mmname{ax-1}))))$, with $|d| = 10$. The
  minimal DAG compression of $d$ is $G_1 = \la \fp_1\pro \D(\mmname{ax-1},
  \mmname{ax-1}),\; \Start$ $\pro \D(\mmname{ax-1}, \D(\mmname{ax-1}, \D(\fp_1,
  \fp_1)))\ra$, with $|G_1| = 2 + 6 = 8$. A stronger compression with
  parameters is $G_2 =
  \la\fp_1(V_1)\pro \D(\mmname{ax-1}, V_1),\;
  \fp_2\pro \fp_1(\mmname{ax-1}),\; 
  \Start\pro \fp_1(\fp_1(\D(\fp_2, \fp_2)))\ra$,
  with $|G_2| = 2 + 1 + 4 = 7$.
\end{examp}
\vspace{-3pt}

Can the MGT of a proof term that is represented by a proof grammar be
determined directly from the grammar? With the \name{grammar-MGT}, defined
below, this is achieved by successively enriching the presupposition base with
MGTs of RHSs. We say that a presupposition base $\base$ is \defname{for a grammar}
$G$ if $\base$ contains for all terminals $p$ with arity~$n$ a
presupposition-statement with name $p$ and arity~$n$.\hspace{-1cm}

\vspace{-3pt}
\begin{defn}[Grammar-MGT]
  \label{def-grammar-mgt}
  Let $G = \la P_1, \ldots, P_n\ra$ be a proof grammar and let $\base$ be a
  presupposition base for $G$. For productions $P_i = p_i(\Vs_i) \pro
  d_i[\Vs_i]$, define the \defname{grammar-MGT} of $p_i(\Vs_i)$ inductively as follows.
  \[\begin{array}{lcl}
  \grammarmgt_{\base,G}(p_i(\Vs_i)) & \eqdef &
  \mgt_{\base'}(d_i[\Vs_i]]),
  \text{ where}\\
  && \base' = \base \cup \bigcup_{j = 1}^{i-1} \{p_j ::
  \grammarmgt_{\base,G}(p_j(\Vs_j))\}.
  \end{array}\]
  In case an involved MGT is undefined, we say that
  $\grammarmgt_{\base,G}(p_i(\Vs_i))$ is \emph{undefined} for all $i \in
  \{1,\ldots,n\}$.\footnote{This criterion could be refined by considering 
  dependencies between nonterminals.}
\end{defn}
\vspace{-3pt}

\enlargethispage{12pt}

The subtlety concerning nonlinear proof terms and the MGT, observed above with
Prop.~\ref{prop-inst}, transfers to the grammar-MGT. Assume
$\grammarmgt_{\base,G}(p_i(\Vs_i))$ is defined. Then
$\mgt_\base(\val_G(p_i(\Vs_i)))$ is defined and, if $G$ is linear,
then \[\grammarmgt_{\base,G}(p_i(\Vs_i)) = \mgt_\base(\val_G(p_i(\Vs_i))).\]
However, in the general case, also for nonlinear $G$, it just holds
that \[\grammarmgt_{\base,G}(p_i(\Vs_i)) \instanceOf
\mgt_\base(\val_G(p_i(\Vs_i))).\]

The sequence of proofs in a \MM database can be viewed as a proof grammar.
However, the database's verified theorem statements, formulated by humans, are
not necessarily the grammar-MGTs of the productions but may be \emph{strict}
instances of these. A \name{KB} (\name{mathematical knowledge base}) thus
explicitly incorporates given theorem formulas:

\vspace{-4pt}
\begin{defn}[KB, Shallow MGT]
  \label{def-kb}
  A \defname{KB} is a triple $K = \la \base, \clauses, G\ra$, where $G = \la
  P_1,\ldots,P_k\ra$ is a proof grammar, $\clauses = \la F_1,\ldots,F_k\ra$ is
  a sequence of definite clauses, and $\base$ is a presupposition base for $G$
  such that for all $P_i = p_i(\Vs_i) \pro d_i[\Vs_i]$ it holds that
  \vspace{-8pt}
  \[F_{i} \instanceOf \shallowmgt_K(p_i(\Vs_i)),\vspace{3pt}\]
  where the \defname{shallow-MGT} of $p_i(\Vs_i)$ is defined as
  \[\begin{array}{lcl}
  \shallowmgt_{K}(p_i(\Vs_i)) & \eqdef &
  \mgt_{\base'}(d_i[\Vs_i]), \text{ where } \base' = \base \cup
  \bigcup_{j = 1}^{i-1} \{p_j :: F_j\}.
  \end{array}\]    
\end{defn}

\noindent
$F_i$ in the definition of KB is a clause whose body length is the arity of
$p_i$. The three considered variations of the MGT are related by
\[F_i \instanceOf \shallowmgt_{K}(p_i(\Vs_i))
  \instanceOf \grammarmgt_{\base,G}(p_i(\Vs_i)) \instanceOf
  \mgt_{\base}(\val_G(p_i(\Vs_i))).\]

\enlargethispage{18pt}
\section{The Core of \setmm as KB}
\label{sec-properties-kbs}

\begin{table}[t]
  \centering
  \caption{Properties of the KB \SETCORE.}
  \label{tab-set-structural}

  \vspace{-3pt}
  \setlength{\tabcolsep}{2pt}    

  \begin{tabular}{rrrrrrrrrrrrrr}\toprule
  && \multicolumn{6}{c}{$\refcount_G(p)$} & \multicolumn{4}{c}{$|p|$} & \multicolumn{2}{c}{$|\val_G(p)|$}\\
  \cmidrule(lr){3-8}\cmidrule{9-12}\cmidrule(lr){13-14}
  \hc{$|G|$} & \hc{$\N(G)$} & \hcmed & \hcavg & \hcmax & \hc{0} & \hc{1} & \hc{$\UpSigma$}
  & \hcmin & \hcmed & \hcavg & \hcmax & \hcmed & \hcmax\\\midrule
   1,824,835 & 27,233 & \ATT 3 & 53 & 63,198 & 16\% & 20\% & 1,444,375 & \ATT 0 & 12 & 67 & 21,651 & \ATT 3$\times 10^{54}$ & 5$\times 10^{1880}$\\\bottomrule
\end{tabular}

\smallskip

\begin{tabular}{rrrrrrrrrrrrrrrrrrrrrr}\toprule
  \multicolumn{6}{c}{$\savval_G(p)$} & \multicolumn{3}{c}{$\arity(p)$} && \multicolumn{4}{c}{$\varoccs(v)$} &
  \multicolumn{3}{c}{$\argmult_G(v)$}\\%
  \cmidrule(lr){1-6}\cmidrule(lr){7-9}\cmidrule(lr){11-14}\cmidrule(lr){15-17}\cmidrule(lr){18-20}
 \hcmin & \hcmed & \hcavg & \hcmax & \hc{${<}0$} & \hc{$0$} & \hcavg & \hcmax & \hc{$0$} & \hc{$\nonlin_G$} & \hcmin & \hcmed & \hcavg & \hcmax & \hcmin & \hcmed & \hcmax\\\midrule
  \ATT -366 & 33 & 796 & 3,981,585 & 12\% & 10\% & \ATT 2 & 28 & 45\% & 28\% & \ATT 0 & 1 & 7 & 2,445 & \ATT 0 & 16,640 & 7$\times 10^{1795}$\\\bottomrule
\end{tabular}

\smallskip

  \begin{tabular}{rrrrrrrrrrr}\toprule
    \multicolumn{4}{c}{$|F|$} & \multicolumn{4}{c}{$\height(F)$} \\
    \cmidrule(lr){1-4}\cmidrule(lr){5-8}
    \hcmin & \hcmed & \hcavg & \hcmax & \hcmin & \hcmed & \hcavg & \hcmax & \hc{${\strictInstanceOf}\f{mgt}$} &
    \hc{$\variant$} & \hc{$\instanceOf$}\\\midrule
    \ATT 0 & 10 & 14 & 193 & \ATT 0 & 4 & 4 & 20 & 8.38\% & 3.08\% & 3.91\%\\\bottomrule
  \end{tabular}

\end{table}

We define $\SETCORE = \la \base, \clauses, G\ra$, a KB that represents the
first 60\% of \setmm. It starts with propositional and predicate calculus,
moves on to set theory and then develops a dozen mathematical topics. The
dropped remainder of \setmm contains guiding and deprecated material as well
as 70 modules of user contributions.
\SETCORE contains 27,233 theorems.\footnote{We exclude the syntactic ``theorems'' with
typecode \mmname{wff}: \mmname{weq}, \mmname{wel}, \mmname{bj-0}, \mmname{sn-wcdeq}.} Its $\base$
component represents the axioms in the considered part of \setmm, $\clauses$
the theorem statements in original order, and $G$ has for each theorem in
$\clauses$ a production representing its proof in \setmm.
Table~\ref{tab-set-structural} shows properties of \SETCORE,
some relating to a specific \emph{network} \cite{newman:2018} extracted from
$G$:

\vspace{-3pt}
\begin{defn}[Proof Dependency Network ($\pdnet$)]
  \label{def-pdnet}
  Let $G$ be a proof grammar. The \defname{proof dependency network} of $G$,
  in symbols $\pdnet(G)$, is defined as the directed graph whose set of nodes
  is the set of all nonterminals of $G$ and whose edges $p \graphedge q$ each
  represent an occurrence of $q$ in the RHS of the production in $G$ that has
  $p$ as nonterminal of its LHS.
\end{defn}

An edge $p \graphedge q$ may be read as ``$q$ occurs as a direct premise of
$p$''.
We now specify the properties shown in Table~\ref{tab-set-structural}, where
$G$ refers to the proof grammar, $p$ to a nonterminal in $G$, and $F$ to a
definite clause in $\clauses$. We use $p$ liberally also to stand for its
production or the LHS of its production. Large multisets of values are
represented by subcolumns with aggregated values. Unless specially noted,
these relate to the \emph{multi}set of values for the members of $G$ or
of $\clauses$.

\begin{coldesc}{1.4cm}\small

  \xitem{$|G|$} Size of $G$, as defined in
  Sect.~\ref{sec-grammar-compressed-trees}.

  \xitem{$\N(G)$} Number of productions of $G$, i.e., number of nodes of
  $\pdnet(G)$.

  \xitem{$\refcount_G(p)$} Number of occurrences of $p$ in RHSs of $G$, i.e.,
  in-degree of $p$ in $\pdnet(G)$. The sum $\UpSigma$ over all members is the
  number of edges of $\pdnet(G)$.

  \xitem{\xmin,\xmed,\xavg,\xmax} Minimum, rounded median, rounded
  average, maximum.

  \xitem{${<}0$, $0$, $1$} Percentage with negative value, value $0$, value
  $1$.

  \xitem{$|p|$} Size of the production of $p$, as defined in
  Sect.~\ref{sec-grammar-compressed-trees}.

  \xitem{$|\val_G(p)|$} Size of $\val_G(p)$, as defined in
  Sect.~\ref{sec-grammar-compressed-trees}.
  
  \xitem{$\savval_G(p)$} Save-value of $p$ in $G$, defined as follows. Let
  $G'$ be $G$ after removing $p$'s production $P$ and unfolding $P$ in all
  RHSs. Then $\savval_G(p)$ is $|G'| - |G|$. The save-value indicates the
  grammar size reduction contributed by $P$. It is 0 if the size remains
  unchanged and negative if the size is increased. For a linear production
  $p(\Vs) \pro d$ it is $\refcount_G(p) * (|d| - \arity(p)) -
  |d|$ \cite{lohrey:treerepair:2013}.
  Presented subcolumns relate to the multiset of the values for just those $p$
  with $\refcount(p) > 0$.

  \xitem{$\arity(p)$} Arity of $p$.

  \xitem{$\nonlin_G$} Percentage of productions of $G$ that are nonlinear.

  \xitem{$\varoccs(v)$} Number of occurrences of variable $v$ in RHSs of $G$.
  Subcolumns refer to the set of all variables $v$ that occur in LHSs of
  productions, assuming that different productions do not share variables.

  \xitem{$\argmult(v)$} Number of occurrences of variable $v$ in $\val_G(p)$
  for the production~$p$ that has $v$ in its LHS. Subcolumns refer to the same
  set of variables as for~$\varoccs(v)$.

  \xitem{$|F|$} Size of $F$. Defined as $|A_0 \revimp_1 A_1 \land \ldots \land
  A_n| \eqdef \UpSigma_{i=0}^n |A_i|$, where $|A_i|$ is defined by $|t| \eqdef
  0$, if $t$ a variable or constant; $|f(t_1,\ldots,t_n)| \eqdef 1 +
  \UpSigma_{i=1}^n |t_i|$, if $n \geq 1$.
  
  \xitem{$\height(F)$} Height of $F$. Defined as $\height(A_0 \revimp A_1
  \land \ldots \land A_n) \eqdef \f{max}(\height(A_0)), \ldots,
  \height(A_n))$, where $\height(A_i)$ is defined by $\height(t) \eqdef 0$, if
  $t$ a variable or constant; $\height(f(t_1,\ldots,t_n))$ $\eqdef 1 +
  \f{max}(\height(t_1), \ldots, \height(t_n))$, if $n \geq 1$.
  
  \xitem{${\strictInstanceOf}\f{mgt}$} Percentage of clauses in $\clauses$
  that are a \emph{strict} instance of the shallow-MGT of the LHS of the
  corresponding production in $G$.

  \xitem{$\variant$} Percentage of clauses in $\clauses$ that would be removed
  if duplicates were deleted such that only a single copy of each member of
  $\clauses$ is retained. Duplicates are considered modulo renaming of
  variables and clause body permutations.

  \xitem{$\instanceOf$} Percentage of clauses in $C$ that would be removed if
  all members that are strictly subsumed by another member were deleted and in
  the resulting multiset duplicates were deleted as described for $\variant$.
  
\end{coldesc}

Table~\ref{tab-set-structural} leads to the following observations and
questions.

\begin{obslist}

  \item \label{obs-save-zero} Concerning $\refcount_G(p)$ and $\savval_G(p)$.
    20\% of the productions are referenced just once. 10\% have save-value 0.
    It seems that these productions have the purpose to break apart a larger
    proof. Do they have further features that may be used to guide breaking
    apart machine-generated proofs, e.g., for presentation?

  \item \label{obs-save-neg} Concerning $\savval_G(p)$. 12\% of the
    productions have a \emph{negative} save-value, although their nonterminal
    is referenced. What is their merit, their role?

  \item Concerning $\varoccs(v)$. Although the median is 1, some productions
    have variables with about 2,500 occurrences. Do these productions play
    special roles?

  \item Concerning $\varoccs(v)$. The minimum 0 indicates productions with a
    variable that does not occur in the RHS, which is so far not even
    considered in our formalization in Sect.~\ref{sec-formal}. Do such
    LHS-only variables have some merit?

  \item Concerning $\arity(p)$. Although the maximum is 28, the average is
    just~2, where 45\% of the productions have arity 0 as in a DAG
    compression.

  \item Concerning $|\protect\val_G(p)|$. Already the median is gigantic,
    indicating inadequacy of ATP techniques that generate just trees.

\item \label{obs-formula-size} Concerning $|F|$ and $\height(F)$. The large
  differences between average and maximum indicate that there are few theorems
  with very large formulas. Do these have typical roles or further features?

\item Concerning ${\strictInstanceOf}\f{mgt}$. With about 8\%, the portion of
  theorems that are \emph{strict} instances of the shallow-MGT and thus strict
  instances of the formula justified by their proof in $G$ (and thus also in
  \setmm) is significant.

\item Concerning $\variant$ and $\instanceOf$. This indicates that about 4\%
  of the theorems of \SETCORE could be removed and replaced by some other theorem
  from \SETCORE. However, this redundancy might have reasons: a statement may
  appear under different theorem names in different application contexts, and
  a strictly subsumed theorem might have a shorter or preferable proof.

\end{obslist}

\section{Compressing Proof Terms}
\label{sec-comparing-mini}

We consider the setting where a set of large ground proof terms is given,
built from constants, $\D$ and $\G$, and we want to compress these into a KB.
After describing techniques for tree compression and reduction, we compare
their result for a small subset of \setmm with the human-expert structuring in
\setmm.

\subsubsection{TreeRePair \cite{lohrey:treerepair:2013}.}
\name{Re-Pair} \cite{repair} is a grammar-based string compression algorithm,
which recursively replaces a most frequent \name{digram}, i.e., pair $fg$ of
consecutive symbols, with a fresh nonterminal $h$, defined with the production
$h \pro fg$. \name{TreeRePair} \cite{lohrey:treerepair:2013} adapts this to
trees. A \defname{digram} is then a pattern
$f(V_1,\ldots,V_{i-1},$ $g(V_i,\ldots,V_{i+m}),V_{i+m+1},\ldots,V_{n-1+m})$
characterized by a parent function symbol~$f$ with arity $n \geq 1$, a child
function symbol~$g$ with arity $m \geq 0$ and an argument index $i$. It has
$n-1+m$ distinct parameters. The corresponding fresh production with
fresh nonterminal $h$ is
\[ h(V_1,\ldots,V_{n-1+m})
    \pro
    f(V_1,\ldots,V_{i-1},g(V_i,\ldots,V_{i+m}),V_{i+m+1},\ldots,V_{n-1+m}). \]
For example, in $\ff(\fg(\fe,\fe),\ff(\fg(\fe,\fe),\fe))$ the digrams with
multiple occurrences are $\ff(\fg(V_1,V_2),V_3)$, $\fg(\fe,V_1)$ and
$\fg(V_1,\fe)$.
TreeRePair operates in two phases, \name{replacement} and \name{pruning}. The
replacement phase proceeds in a loop by modifying a main term initialized with
the input term. In each round it identifies digrams with multiple occurrences,
selects one or more according to heuristic criteria, generates a fresh
production for each, and, in the main term, ``folds into'' these productions,
i.e., rewrites from RHS to LHS. Since digrams may overlap, this can be
concretized in different ways.
Our Prolog implementation maintains the main term as DAG and supports various
options. The output of this phase is a grammar with the fresh productions
and a start production with the final stage of the main term as RHS. The
grammar is a \name{proof grammar} as specified in
Sect.~\ref{sec-grammar-compressed-trees}.
In the pruning phase, productions whose save-value is negative or zero are
eliminated by unfolding them in all RHSs. Since the result depends on how
these productions are selected \cite{lohrey:treerepair:2013}, this is configurable
in our implementation.

\subsubsection{Nonlinear Compression.}

TreeRePair generates only \emph{linear} grammars. Nonlinear grammars achieve a
stronger compression and about one third of the proofs of \setmm are
nonlinear. \defname{Nonlinear compression} converts a given proof grammar to a
possibly nonlinear one. It modifies the grammar in a loop where in each round
a production $P = p(\Vs) \pro d[\Vs]$ is chosen such that in some RHS there is
an instance of $p(\Vs)$ with two identical arguments. A fresh nonlinear
production is then generated, whose LHS $p'(\Vs')$ has one argument less than
$p$ and whose RHS is $d[\Vs]$ with the two parameters corresponding to
repeated arguments identified. In the RHSs of the grammar then all instances
of $p(\Vs)$ where the respective arguments are identical are rewritten to
$p'(\Vs')$, and the grammar is pruned, as by TreeRePair. The final outcome is
sensitive to the choice of $P$ in each round. In experiments we picked $P$
with lowest save-value.
As we have seen in Sect.~\ref{sec-grammar-compressed-trees}, with a nonlinear
grammar for an expanded proof term with a defined MGT, the
grammar-MGT may be a \emph{strict} instance or even be undefined. For
nonlinear compression we thus provide options to pick only $P$ where specified
properties of grammar-MGTs are preserved. In particular, that grammar-MGTs are
defined and that for given nonterminals intended as top-level theorems the
grammar-MGT subsumes a given theorem formula.

\subsubsection{Same-Value Reduction.}
Grammars obtained by TreeRePair and nonlinear compression can have multiple
nonterminals with the same expanded value. %
\defname{Same-value reduction} replaces these by a single one.

\subsubsection{MGT-based Reduction.}
The grammar-MGTs for two nonterminals can be variants, or one can strictly
subsume the other, possibly modulo clause body permutation. \name{MGT-based
  reduction} eliminates these cases by replacing redundant nonterminals in the
RHSs and removing their productions. Our implementation allows to parameterize
this, and also same-value reduction, such that nonterminals intended as
top-level theorems are protected from elimination.

\subsubsection{Experiments with \MiniSet, a Small Subset of \setmm.}

The shown compression techniques seem not (yet) suitable to process all
expanded proof trees of \SETCORE. To find a feasible subset, we looked at
\name{Theorem Sampler} \cite{metamath:theoremsampler}, which highlights 44
theorems from \setmm, and selected those 17 that could be fully expanded and
compressed in under a minute: \mmname{idALT}, \mmname{peirce},
\mmname{anim12}, \mmname{exmid}, \mmname{pm5.18}, \mmname{consensus},
\mmname{meredith}, \mmname{19.12}, \mmname{19.35}, \mmname{equid},
\mmname{2eu5}, \mmname{uncom}, \mmname{abeq2}, \mmname{isset}, \mmname{ru},
\mmname{peano3}, \mmname{ac2}. The median processing time was 0.5~s.
The human-expert structuring in \setmm is obtained by supplementing
productions for the proofs of all theorems that are directly or indirectly
referenced in the proofs of these 17 top-level theorems. We call the resulting
KB \MiniSet.
Then we took the set of the fully expanded proof terms for the 17 top-level
theorems, where the sum of their sizes is 5$\times 10^{22}$, and applied our
compression methods. We call the resulting KB \MiniTrp. The replacement phase
of TreeRePair yields a grammar of \name{grammar size}/\name{number of
  productions} 9,739/4,153. Pruning yields 3,683/905. Nonlinear compression
3,204/604. Same-value reduction 3,174/593 and, finally, MGT-based reduction
3,017/534. The $\clauses$ component of \MiniTrp contains for the 17 top-level
theorems the original theorem formulas, which were kept protected in the
compressions. Table~\ref{tab-miniset-structural} compares \MiniSet and
\MiniTrp. It also shows values of \MiniDag, which is the minimal DAG grammar
obtained from the same input as \MiniTrp. Compression \MiniXC will be
discussed in Sect.~\ref{sec-humancompress}. On the basis of
Table~\ref{tab-miniset-structural} we observe the following.

\begin{table}[t]
  \centering
  \caption{Properties of \MiniSet, \MiniTrp and further compressions.}
  \label{tab-miniset-structural}
  \vspace{-4pt}
  
    \setlength{\tabcolsep}{2.4pt}

    \begin{tabular}{lrrrrrrrrrrrrrrrr}\toprule
      && & \multicolumn{5}{c}{\textit{$\refcount_G(p)$}} & \multicolumn{4}{c}{\textit{$|p|$}}
      & \multicolumn{2}{c}{$|\val_G(p)|$}\\%
      \cmidrule(lr){4-8}\cmidrule(lr){9-12}\cmidrule(lr){13-14}
      & $|G|$ & $\N(G)$ & \hcmed & \hcavg & \hcmax & 1 & \hc{$\UpSigma$} & \hcmin & \hcmed & \hcavg & \hcmax & \hcmed &
      \hcmax\\\midrule
      \MiniSet & 2,302 & 690 & \ATT 2 & 3 & 68 & 45\% & 2,114 & \ATT 0 & 3 & 3 & 68 & \ATT 46,647 & 5.06$\times 10^{22}$\\
      \MiniTrp & 3,017 & 534 & \ATT 3 & 5 & 85 & 0\% & 2,900 & \ATT 1 & 3 & 6 & 288 & \ATT 11,034 & 1.29$\times 10^{20}$\\
      \MiniDag & 21,472 & 927 & \ATT 2 & 7 & 966 & 0\% & 6,894 & \ATT 0 & 8 & 23 & 1,694 & \ATT 17,171,018 & 5.06$\times 10^{22}$\\
 \MiniXC & 1,831 & 339 & \ATT 3 & 5 & 58 & 2\% & 1,700 & \ATT 1 & 3 & 5 & 165 & \ATT 48,640 & 1.26$\times 10^{23}$\\\bottomrule
    \end{tabular}

    \smallskip

    \begin{tabular}{lrrrrrrrrrrrrrrrrrr}\toprule
      & \multicolumn{6}{c}{$|\savval_G(p)|$} & \multicolumn{3}{c}{$\arity(p)$} && \multicolumn{4}{c}{$\varoccs(v)$} & \multicolumn{4}{c}{$\argmult_G(v)$}\\%
      \cmidrule(lr){2-7}\cmidrule(lr){8-10}\cmidrule(lr){12-15}\cmidrule(lr){16-19}
      & \hcmin & \hcmed & \hcavg & \hcmax & \hc{${<}0$} & \hc{$0$} & \hcavg & \hcmax & 0 & $\nonlin_G$ & \hcmin & \hcmed & \hcavg & \hcmax &
      \hcmin & \hcmed & \hcavg & \hcmax\\\midrule
      \MiniSet & -5 & 0 & 3 & 358 & 31\% & 29\% & \ATT 1 & 5 & 40\% & 2.17\% & \ATT 1 & 1 & 1 & 2 & \ATT 1 & 1 & 37 & 18,432\\
      \MiniTrp & \ATT 0 & 4 & 25 & 7,063 & 0\% & 0\% & \ATT 1 & 7 & 48\% & 2.43\% & \ATT 1 & 1 & 1 & 3 & \ATT 1 & 1 & 1 & 4\\
      \MiniDag & \ATT 1 & 18 & 71 & 16,140 & 0\% & 0\% & \ATT 0 & 0 & 100\% & 0.00\% & \ATT \ATT \\
      \MiniXC & \ATT -2 & 4 & 13 & 1,451 & 1\% & 2\% & \ATT 1 & 3 & 45\% & 2.06\% & \ATT 1 & 1 & 1 & 2 & \ATT 1 & 1 & 6 & 512\\\bottomrule
    \end{tabular}

    \smallskip    
    \setlength{\tabcolsep}{1.8pt}
    \begin{tabular}{lrrrrrrrrrrrrrrrrrr}\toprule
      & \multicolumn{4}{c}{$|F|$} & \multicolumn{4}{c}{$\height(F)$}\\
      \cmidrule(lr){2-5}\cmidrule(lr){6-9}
      & \hcmin & \hcmed & \hcavg & \hcmax & \hcmin & \hcmed & \hcavg & \hcmax &
      \hc{$\xset$} &
      \hc{$\xmm$}\\\midrule
      \MiniSet & \ATT 0 & 5 & 6 & 48 & \ATT 0 & 3 & 3 & 13\\
      \MiniTrp & \ATT 0 & 6 & 7 & 74 & \ATT 0 & 3 & 3 & 11 & \ATT 34\% & 29\%\\ %
      \MiniDag & \ATT 0 & 7 & 10 & 53 & \ATT 0 & 4 & 5 & 13 & \ATT 21\% & 18\%\\ %
      \MiniXC &  \ATT 0 & 5 & 5 & 23 & \ATT 0 & 3 & 3 & 10\\\bottomrule
    \end{tabular}
    
\end{table}

\begin{obslist}
  
\item Concerning $|G|$. \MiniTrp is about 30\% larger than \MiniSet. What
  mechanical techniques are missing for a comparable compression rate?
  
\item Concerning $|G|$ and $\val_G(p)$. DAG compression already brings the
  gigantic tree sizes down to feasibility for machines. Pattern-based grammar
  compression reduces the size further by a factor of about 7--10.

\item Concerning $\savval_G(p)$. Noticeably larger in \MiniTrp  than in \MiniSet.

\item Concerning $\argmult(p)$. The maximum in \MiniSet is strikingly large,
  whereas the median is similar to \MiniTrp.

\item Concerning \xsc{$\xset$} and \xsc{$\xmm$}. These columns indicate the
  amount by which machine compression coincides with structuring by humans.
  34\% of the formulas in \MiniTrp are also in \setmm, and 29\% are also in
  \MiniSet, i.e., rediscoveries of lemma formulas from human structuring. The
  17 top-level theorems themselves are not counted there and comparison is
  modulo body permutations. The difference between both values indicates that
  5\% of the formulas of \MiniTrp are in \setmm and potentially useful for
  proving the 17 top-level theorems, but were not used to prove them in the
  human structuring. Possibly some of these come in \setmm only after the
  theorem for which they might have been used.
  
\end{obslist}

\subsubsection{The Network View.}
\begin{figure}[t]
  \centering
  \begin{minipage}[c]{12.55cm}
    \includegraphics[width=0.32\linewidth]{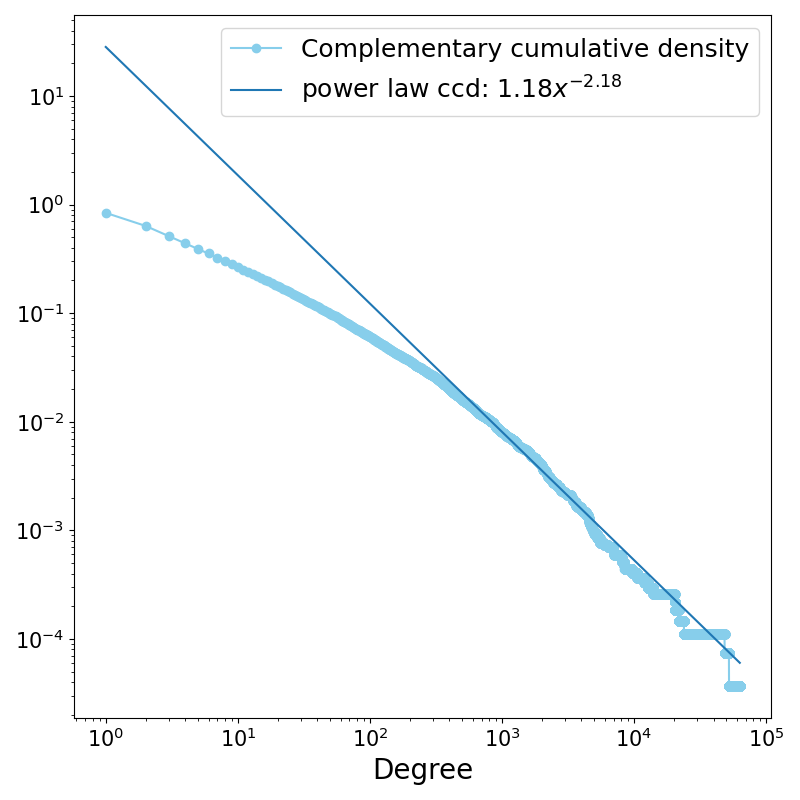}
    \includegraphics[width=0.32\linewidth]{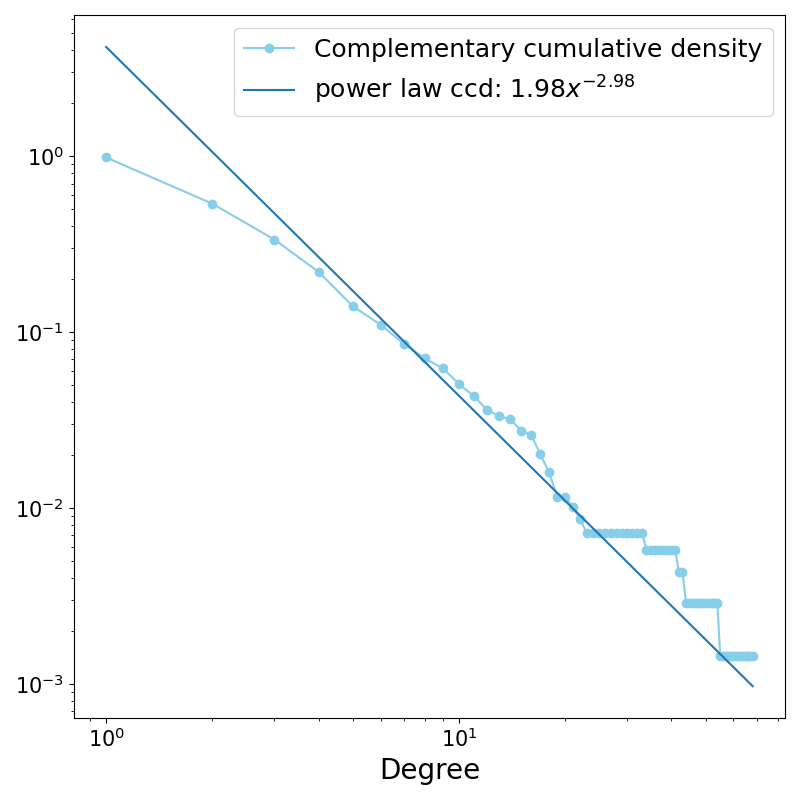}
    \includegraphics[width=0.32\linewidth]{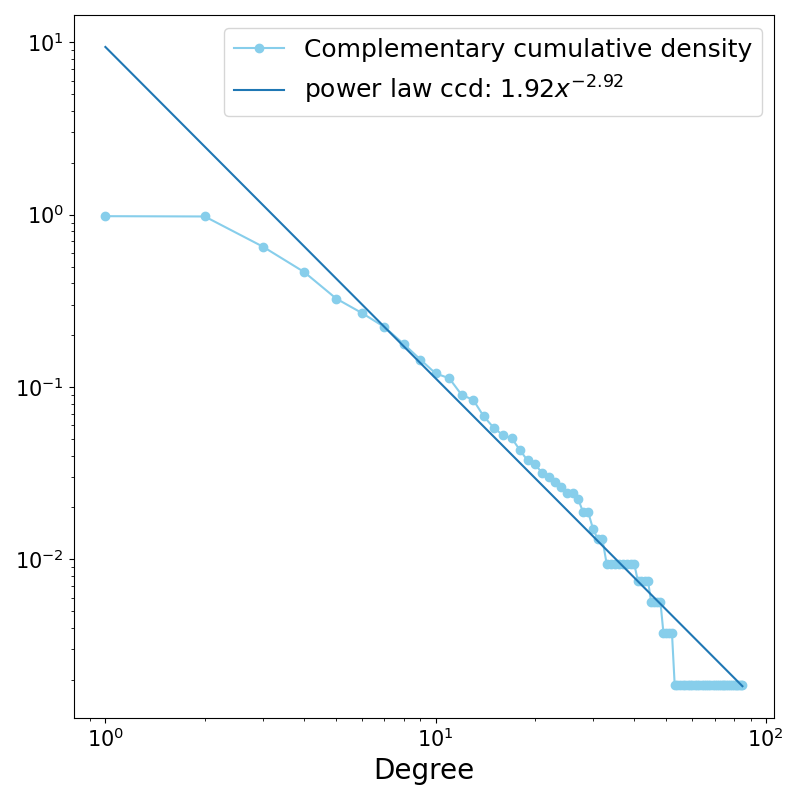}
  \end{minipage}
  \vspace{-14pt}  
  \caption{Complementary cumulative distribution of the empirical in-degree,
    along with that of the fitted scale-free graph. Both axes are of
    logarithmic scale. The fitted curves are shifted to align with the tail of
    the distribution. \textbf{Left:} \SETCORE \textbf{Center:} \MiniSet
    \textbf{Right:} \MiniTrp}
  \label{fig:power_law_setcore_miniset_minitrp}
\end{figure}

The proof dependency network $\pdnet(G)$ (Def.~\ref{def-pdnet}) of the grammar
$G$ of a KB $\la \base, \clauses, G\ra$ provides an abstracted view on the
proof structure that is accessible to the study of complex networks
\cite{newman:2018}. We call it the \defname{\PDNET} of the KB and ask for
characteristics of these networks, as well as for differences depending on
whether they were created by humans or by machine.
Recall that in a \PDNET an edge $p \graphedge q$ expresses that $q$ occurs as
a direct premise of $p$ and that $\refcount_G(p)$ is the in-degree of node $p$
in $\pdnet(G)$. We make an unexpected observation about the in-degree
distributions of the \PDNETs of \SETCORE, \MiniSet and \MiniTrp
(Fig.~\ref{fig:power_law_setcore_miniset_minitrp}):
\begin{obslist}
\item \label{obs-scalefree} \PDNETs seem to have power-law degree distributions, a property also
  known as \name{scale-free}. That is, if we uniformly select a node, then the
  probability of the node having in-degree $k$ is given by a density function
  $p(k) = (\alpha-1) k^{-\alpha}$, where $\alpha$ is a hyperparameter, called
  \emph{scaling factor}. We can fit the scaling factor for a \PDNET according
  to the maximum likelihood estimate \cite{powerlaw}.
    Figure~\ref{fig:power_law_setcore_miniset_minitrp} shows in-degree
    distributions. Both axes are logarithmic and a straight line indicates
    perfect fit to a power law distribution. We observe that the curves
    deviate from power law for small degrees, but the tail fits nicely.
\end{obslist}

\begin{figure}[t]
  \centering
  \vspace{-8pt}
    \includegraphics[width=0.4\linewidth]{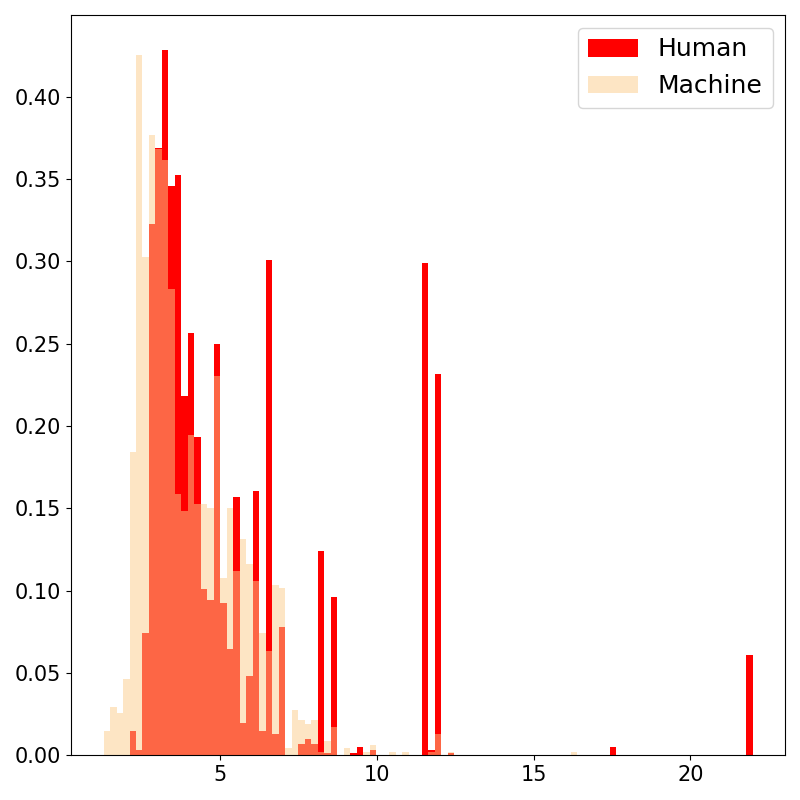}
    \includegraphics[width=0.4\linewidth]{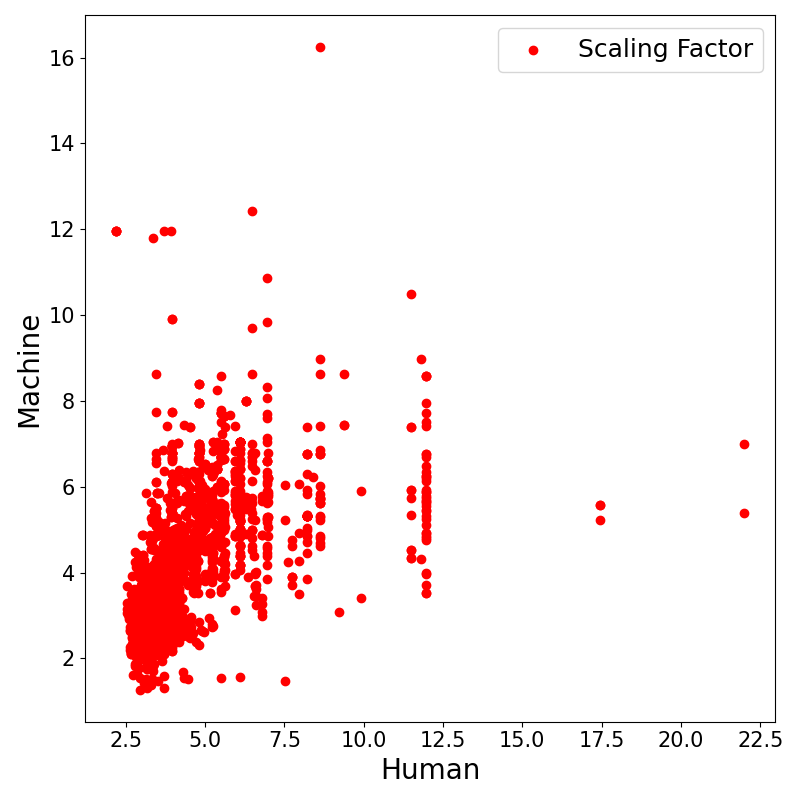}
    \vspace{-12pt}    
  \caption{\PDNETs of proof grammars as found in \MM (Human) and of those
    generated via automated compression from the fully expanded proof tree
    (Machine): \textbf{Left:} Histogram of power law distribution scaling
    factors, fitted to each \PDNET separately.  \textbf{Right:} Scatter plot
    showing the correlation among scaling factors of both \PDNETs.}
  \label{fig:scaling_factor_histogram}
\end{figure}

To compare human and machine structurings,
we select the theorems
from \SETCORE whose proof could be expanded and compressed with the workflow
for \MiniTrp (Sect.~\ref{sec-comparing-mini}) in 100~s. This gives a set of
3,287 pairs $\la$\name{human}$,$\name{machine}$\ra$ of \PDNETs: \name{human}
for the subset of \SETCORE with the theorem itself and the theorems that are,
directly or indirectly, referenced in its proof; \name{machine} for the
compressed expansion. Fig.~\ref{fig:scaling_factor_histogram} shows
correlations among these. We observe the following.

\begin{obslist}
\item The scaling factors of \PDNETs for human structurings and for machine
  structurings are \emph{strongly correlated}, with no clear difference.
 \end{obslist}

Many real-world networks exhibit power law degree distributions
\cite{scalefree,powerlaw,newman:2018}. Often a power law tends to hold only
for the tail of the distribution and not for nodes with smaller degree
\cite{powerlaw}. As far as we are aware, this is the first time that this
phenomenon is identified in graphs of formal proofs. We note, however, that
there is an ongoing debate about how widespread such distributions are
\cite{scalefree_rebuttal}.

\section{Compressing Human Structurings Further}
\label{sec-humancompress}

We investigate ways to \emph{combine} a given human-structured proof grammar with
grammar compressions and reductions by machine.
Our key technique is applying TreeRePair to a given \emph{grammar}, to obtain
a stronger compression:
We take the set of all RHSs of the grammar, combine them into a single term
by introducing a new virtual root node and compress the resulting term with
the workflow described in Sect.~\ref{sec-comparing-mini}.
Combined with nonlinear compression, pruning and grammar reductions, this
reduces \MiniSet from grammar size 2,302 to 1,831. Properties of the resulting
KB \MiniXC are shown in Table~\ref{tab-miniset-structural}.

We now look at larger collections of theorems from \setmm and explore how well
they can be compressed with our technique. We focus on \emph{adding} novel
lemmas, modifying the proofs of the given theorems such that they use these
lemmas.
For each of the 18 topics $\mathcal{T}$ in \SETCORE we consider the grammar
$G_{\mathcal{T}}$ consisting of all productions (i.e., proofs) of the theorems
of the topic.
We apply the described compression to these grammars
$G_{\mathcal{T}}$, configured such that productions for original nonterminals in
$G_{\mathcal{T}}$ are never eliminated.
It turns out that TreeRePair applied to these proof grammars effects a strong
further compression: If we compare \SETCORE with the union of all the
resulting grammars, we obtain a reduction of 7\%. Among the individual topics,
the least compression is observed for \name{Basic Algebraic Structures} (4\%)
and the greatest for \name{Tarski-Grothendieck Set Theory}~(30\%).

While the compression is in part due to a large number of rarely used lemmas,
we also find lemmas in each topic that are heavily used, indicating
that they are
potentially useful and worth considering to be added to the \MM corpus. For
example, in the \name{Axiom of Choice} topic a frequent lemma  is given by
the production
\begin{verbatim}
  lemma905(A) -> ad2antrr(syl(A, necon2ai(mtbii(sdom0, breq2)))).
\end{verbatim}
The proven formula in \MM syntax (with Prolog variables) is
\begin{verbatim}
  $e |- ( A -> B ~< C ) $.
  $p |- ( ( ( A /\ D ) /\ E ) -> (/) =/= C ) $.
\end{verbatim}
A selection of generated lemmas for the 18 topics is shown in App.~\ref{app:human_compress}.

\section{Conclusion}
\label{sec-conclusion}

We conclude with discussing features of our approach and system in the context of
potential applications and related works.

\subsubsection{Structuring Proofs from Automated Systems.}
\label{sec-structuring}

Proofs by automated systems are typically unnecessarily long
and structured ad-hoc.
For presentation and exchange, important steps
should be separated from trivialities, taking into account background
knowledge and terminology of the client.
Schulz \cite{schulz:projektarbeit:1993,denzinger:schulz:1994} presents a
system to structure such proofs, where a proof is represented by a graph -- our
$\pdnet$ but with edges directed conversely. The method awards status
``\name{lemma}''
to nodes with estimated high importance. Among seven investigated criteria,
the three most powerful are structure-based: \name{frequently used steps},
which is our $\refcount_G(p)$; \name{important intermediate results}, which is
$\savval_G(p)$ for DAGs; and \name{isolated proof segments}, which captures
the idea that a lemma is important \emph{for a given proof} if it is used
often within that proof but rarely from outside. We have implemented but not
yet explored that latter idea.

Vyskocil, Stanovský and Urban \cite{vyskocil:stanovsky:urban:definitions:2010}
designed and implemented a method for automated proof compression by
\emph{invention of new definitions}, with goals similar to ours.
Their technique is in essence grammar-based tree compression, however, applied
to \emph{formulas} involved in proofs,
as also studied by Hetzl
\cite{hetzl:tree:2012}, in contrast to the \emph{structure} of proofs.
In \cite{vyskocil:stanovsky:urban:definitions:2010} it is observed that
mechanically obtained definitions are rarely interesting from a
mathematician's point of view, and that learning new definitions is costly for
human readers.
In contrast, we find that compressing proof \emph{structures} tends to yield
interesting lemmas, as indicated by the fraction of rediscovered theorems from
\setmm (Table~\ref{tab-miniset-structural}) and intuitively convincing
discovered lemmas (Sect.~\ref{sec-humancompress}).
Adaptation of mechanical structurings to existing client terminology (theorem
names/nonterminal symbols) is supported in our system through operations for
merging proof grammars.

\subsubsection{Hammering.}
\label{sec-hammering}

\name{Hammer systems} \cite{blanchette:etal:hammering:2016} address the
interplay between automated theorem provers and interactive proof assistants,
which come with large verified mathematical knowledge bases. Typically, a
hammer system invokes (1.)~a premise selector that identifies a fraction of
available theorems as potentially relevant for the current interactive goal,
(2.)~a translation module that constructs an ATP problem from the premises and
the goal, and (3.)~a proof reconstruction module that converts the proof by
the ATP system into a form accepted by the proof assistant.
Carneiro, Brown and Urban \cite{metamath:atp:2023} present a hammer system for
\Metamath. They support different formula translations via \name{Metamath
  Zero} \cite{carneiro:mm0:2020}, into higher-order logic, and from there to
definite first-order clauses. ATP systems can be incorporated and benchmarked
with different translations and premise selection models. \ProverN
\cite{prover9} outputs proofs of first-order Horn problems that are
suitable for proof reconstruction, which, however, involves expanding the
resolution proof DAG into a tree.

So far, we only provide some elements of a hammer system. Our \MM interface is
implemented from scratch in \SWIPL. \MM's native representation of formulas is
as sequences of constant and variable tokens, separated by spaces, e.g.,
\mmname{( ph -> ( ps -> ph ) )}, where parsing into formula terms is done
\emph{within proofs}, based on declarations in the database. It is easy to
strip off these ``syntactic'' parts from proofs. The \mmexe tool
\cite{metamath:book} shows proofs by default without them. We also do not
consider them in proof terms, although our interface makes them available. We
perform formula parsing with Prolog DCG grammars generated on the fly from
relevant database declarations, which leads to the same formulas as the
first-order translation of \cite{metamath:atp:2023}. Our system can use syntax
declarations to pretty-print formulas in \MM notation. To export proofs for
\MM, we have a procedure that \emph{infers} a suitable syntactic part to be
supplemented to a given proof, on the basis of declarations, subject to \MM's
inheritance mechanism of \name{disjoint variable restrictions}
\cite{metamath:book}.

\subsubsection{Premise Selection and Querying Mathematical Knowledge Bases.}

Ka\-li\-szyk and Urban \cite{kaliszyk:urban:millions:2015}
address premise selection. They observe that relevant lemmas are not only
found among the named theorems of a knowledge base corpus, but also among
lemmas used implicitly in proofs. This can be taken into account at levels
with different granularity: Lemmas can be identified at the level of
``atomic'' kernel inferences, which leads to big data, and at the higher level
of combinations of ``tactics''.
Our approach, based on grammar-compressed proof structures, integrates both
levels: on the one hand we have the fully expanded proof trees with gigantic
size, on the other hand the grammar compression that can be verified in
seconds and provides a distinguished lemma formula for each of its
productions. There is no loss of information between the levels. Both
represent the proof in full detail, and both use the same representation
mechanism.
Of course, this is tied to the commitment to a specific proof system, that of
\Metamath, based on condensed detachment. Transfer to other systems remains to
be investigated. A possibility might be via a translator
\cite{carneiro:conversion:2014}. Also, condensed detachment with compression
can simulate ATP calculi, e.g., resolution \cite{cw:ccs:2022}.

In our framework, \MM theorem formulas are available as definite first-order clauses with a
single unary predicate, directly suited for searching all theorems that
subsume a given conjecture. For more advanced forms of queries, e.g.,
\cite{cw:mathlib:97}, they can be used as premises of ATP tasks.

\subsubsection{Analysis of Formal Proofs.}

\MM is intended not just for archiving and verifying mathematical proofs, but
also for \emph{studying} them. Our approach and tools offer here many
possibilities, e.g., determining structural and formula-related proof
properties, for single proofs and for sets of inter-related proofs, and for
proofs by humans, by machine or combinations of both. Redundancies can be
observed and eliminated, or be intentionally introduced. Proof transformations
-- particularly convenient in the term view -- can tailor proofs to
requirements.

\subsubsection{Outlook.}

Although we addressed many issues in the combination of
proof assistants with their large verified knowledge bases and automated
theorem proving, the presented work is in a sense just a start. Observations
made in experiments call for dedicated investigations. The network view may be
elaborated in depth. Practical tools for specific tasks could be derived. A
distinguishing feature of our approach is compression of \emph{proof
structures}, in contrast to formulas. Are there inherent relationships between
them? Combinatory logic \cite{schoenfinkel:24:bausteine,curry:1958} provides
an alternative way to express compressions \cite{cw:ccs:2022}. Does this lead to
interesting results? Can problems with extremely long first-order and short
higher-order proofs \cite{boolos:curious:1987,benzmueller:etal:boolos:2023} be
approached with compression? How do our compressions combine with other proof
reductions, e.g., \cite{cwwb:article:2024}? Are there ways to tightly
integrate ATP systems, notably those based on generating proof structures
\cite{cw:ccs:2022,rwzb:lemmas:2023,cw:sgcd:2024,cwwb:lukas:2021,cwwb:article:2024}?
How can insights from compression help to guide provers?

\nocite{luk:selected:1970}
\bibliographystyle{splncs04}
\bibliography{bibmetamath2025}

\clearpage
\appendix
\section{Additional Plots of Degree Distributions of \PDNETs}
\label{app:power_law_fit}

\begin{figure*}[htb]
  \begin{center}
    \includegraphics[width=0.32\linewidth]{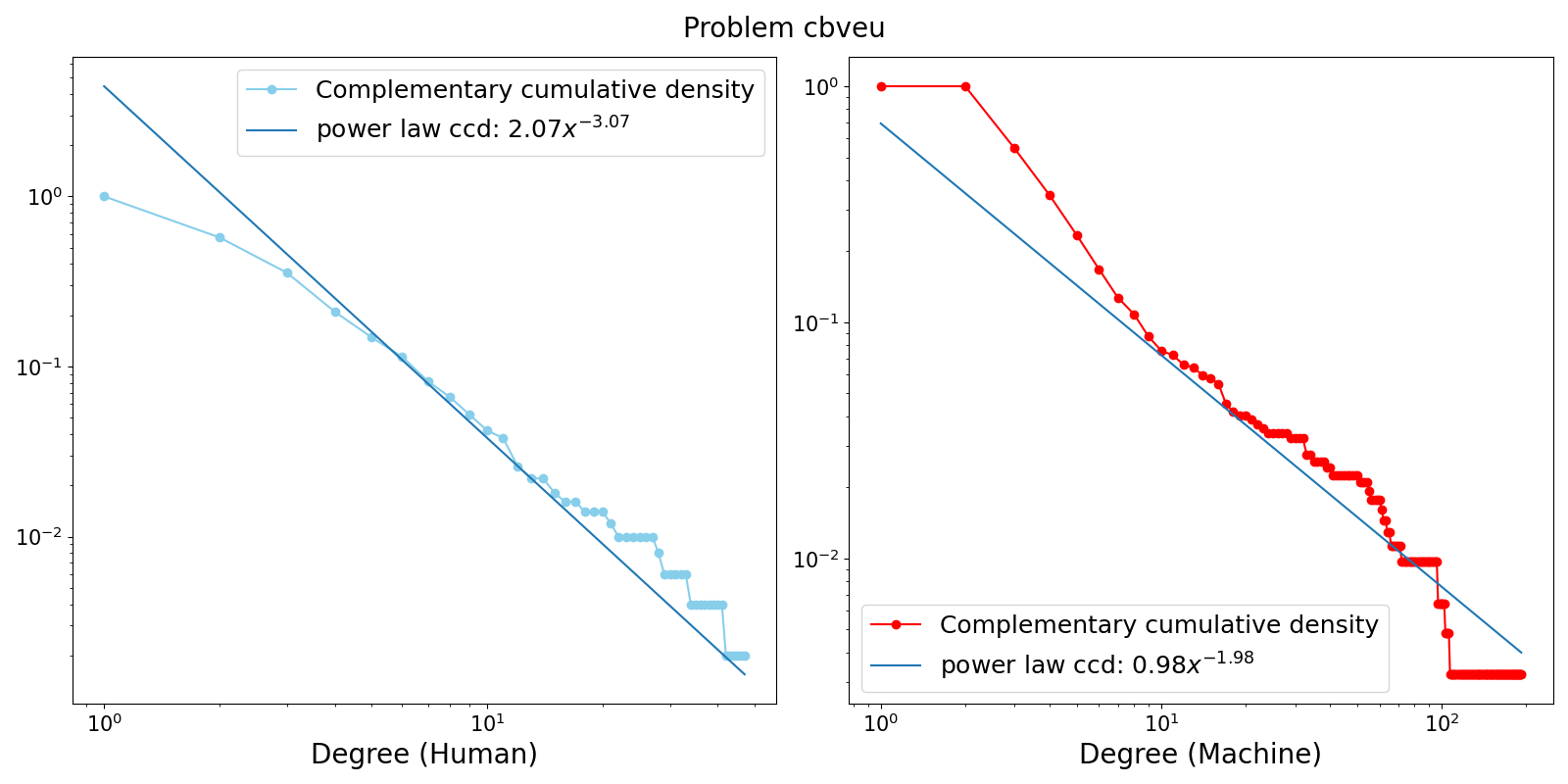}
    \includegraphics[width=0.32\linewidth]{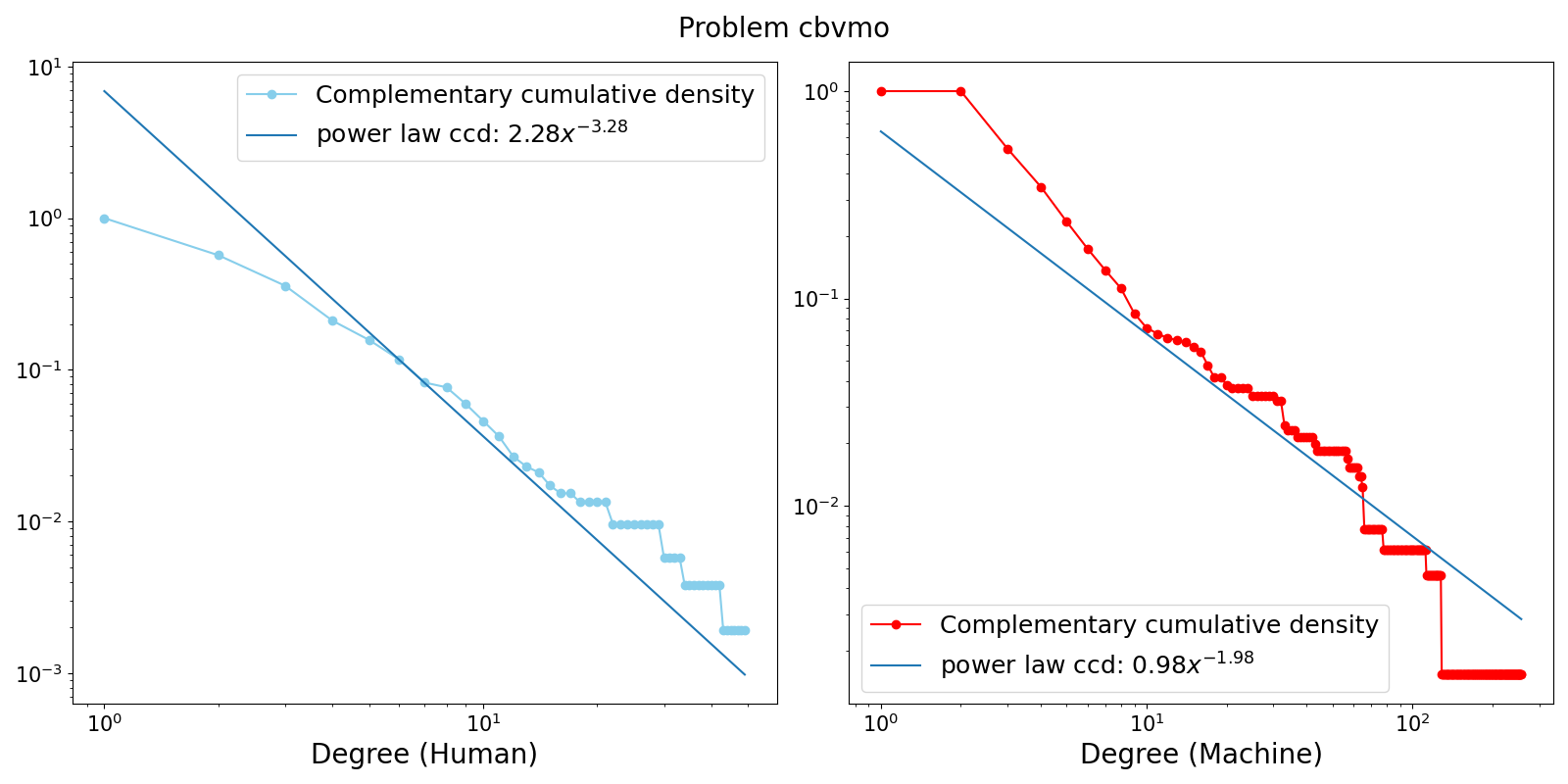}
    \includegraphics[width=0.32\linewidth]{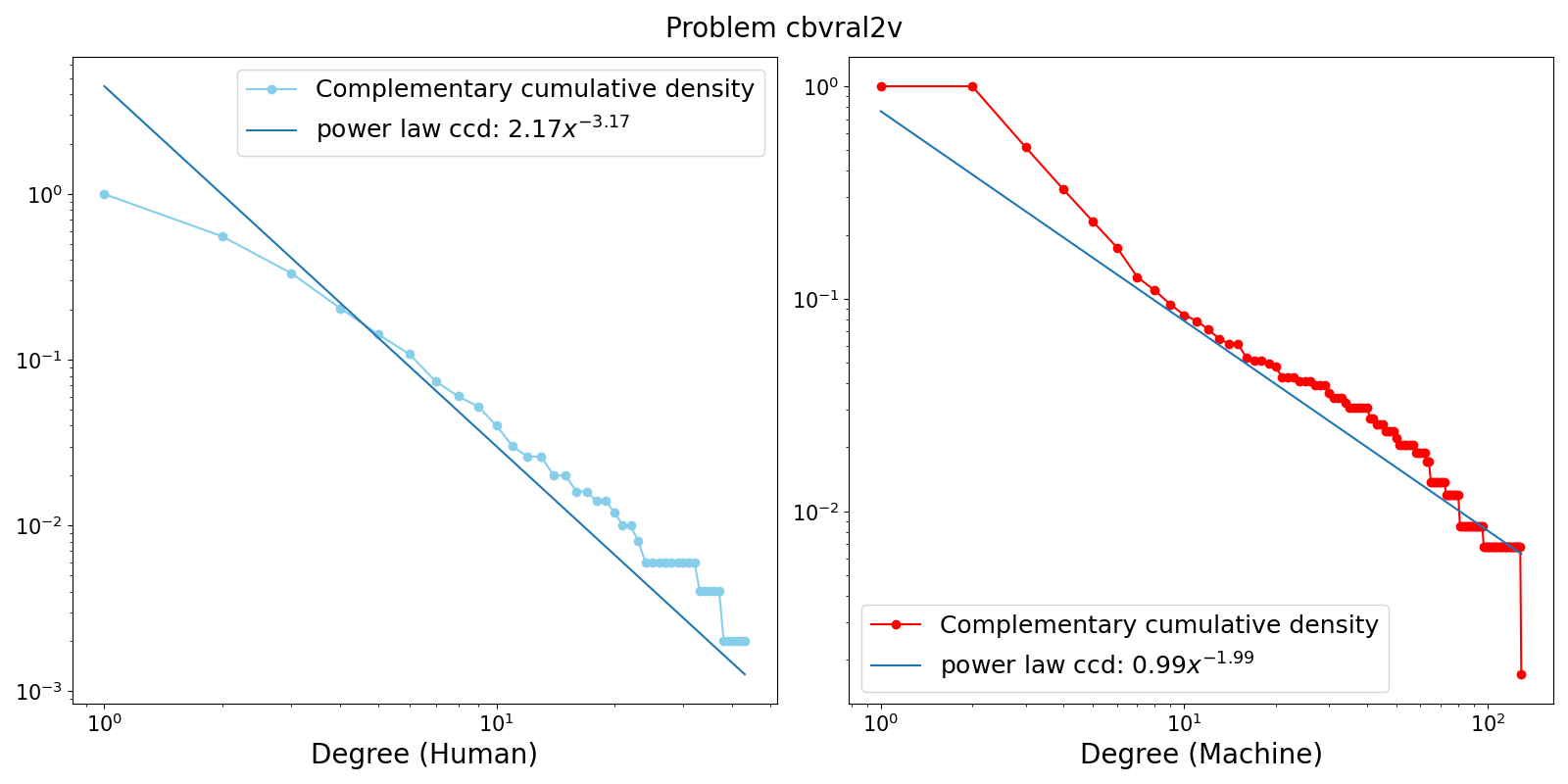}
    \includegraphics[width=0.32\linewidth]{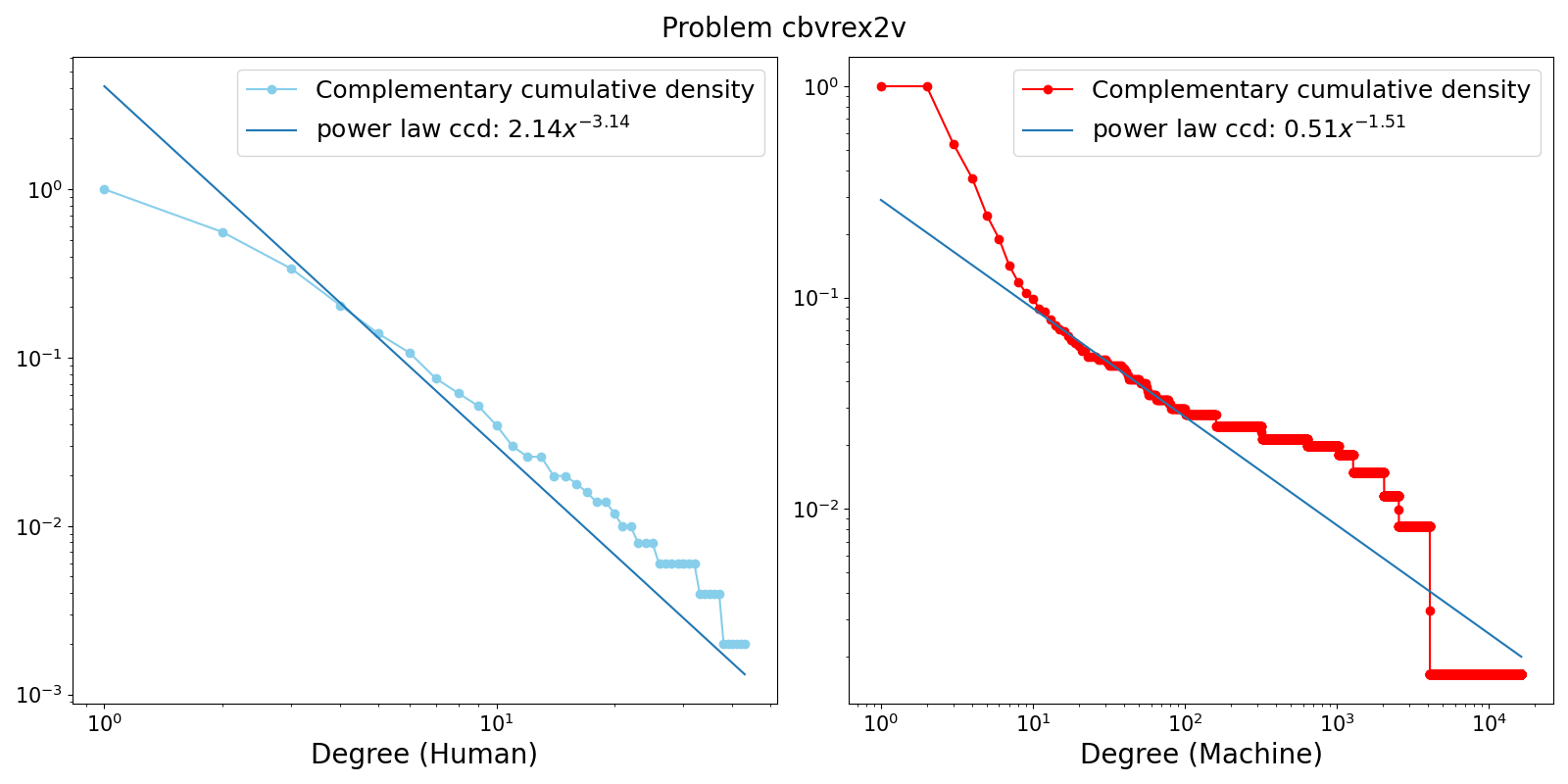}
    \includegraphics[width=0.32\linewidth]{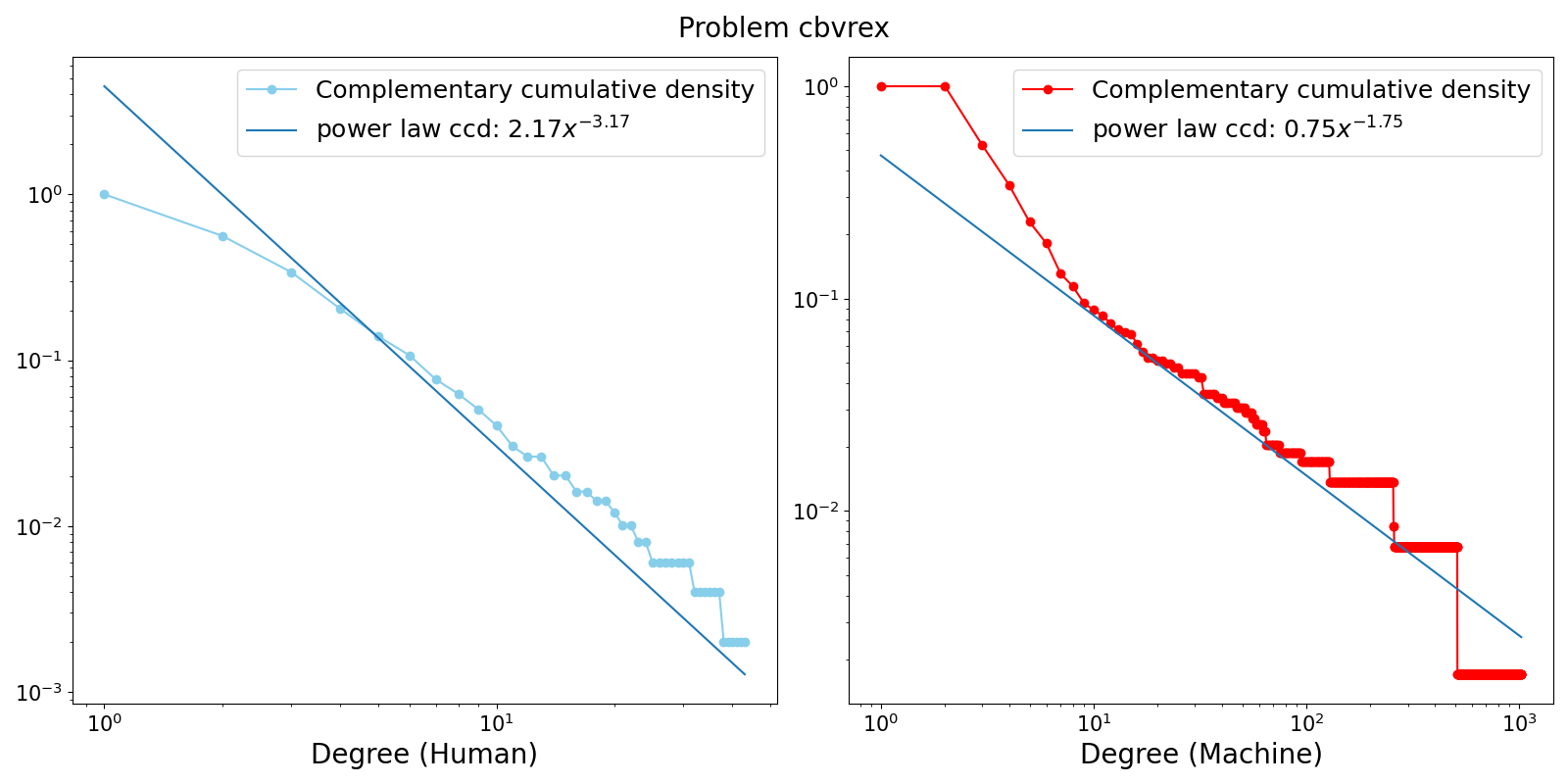}
    \includegraphics[width=0.32\linewidth]{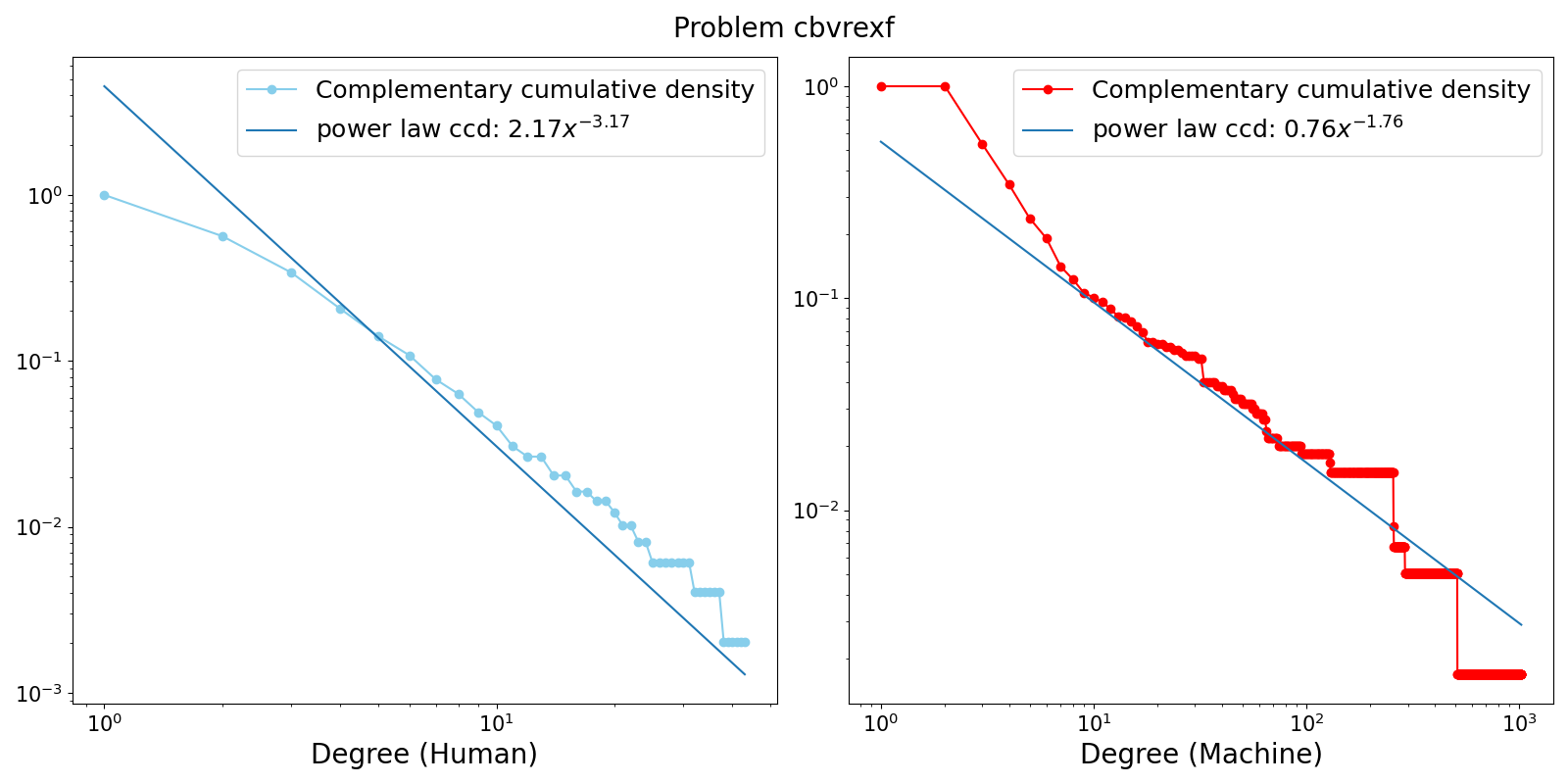}
    \includegraphics[width=0.32\linewidth]{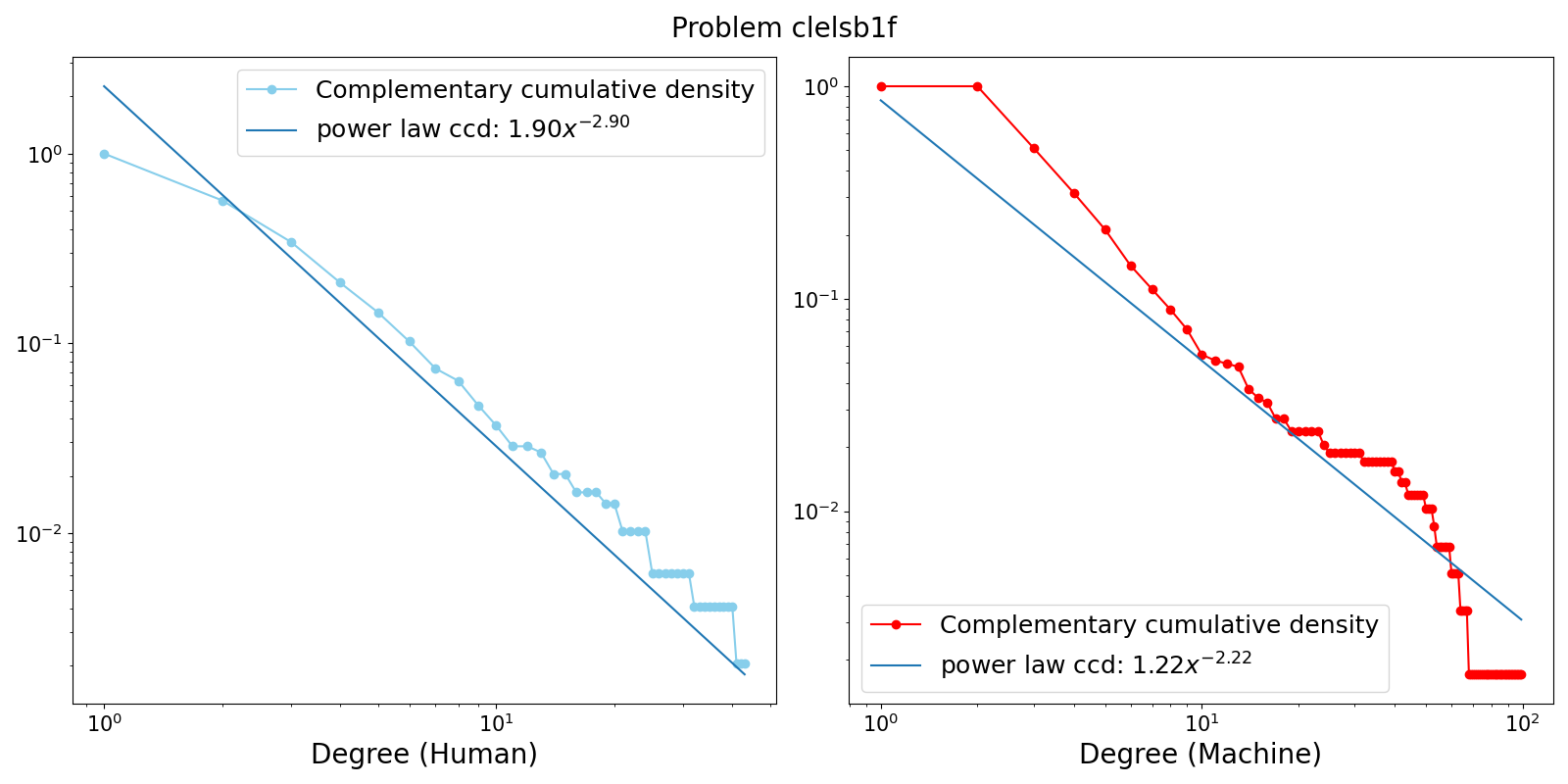}
    \includegraphics[width=0.32\linewidth]{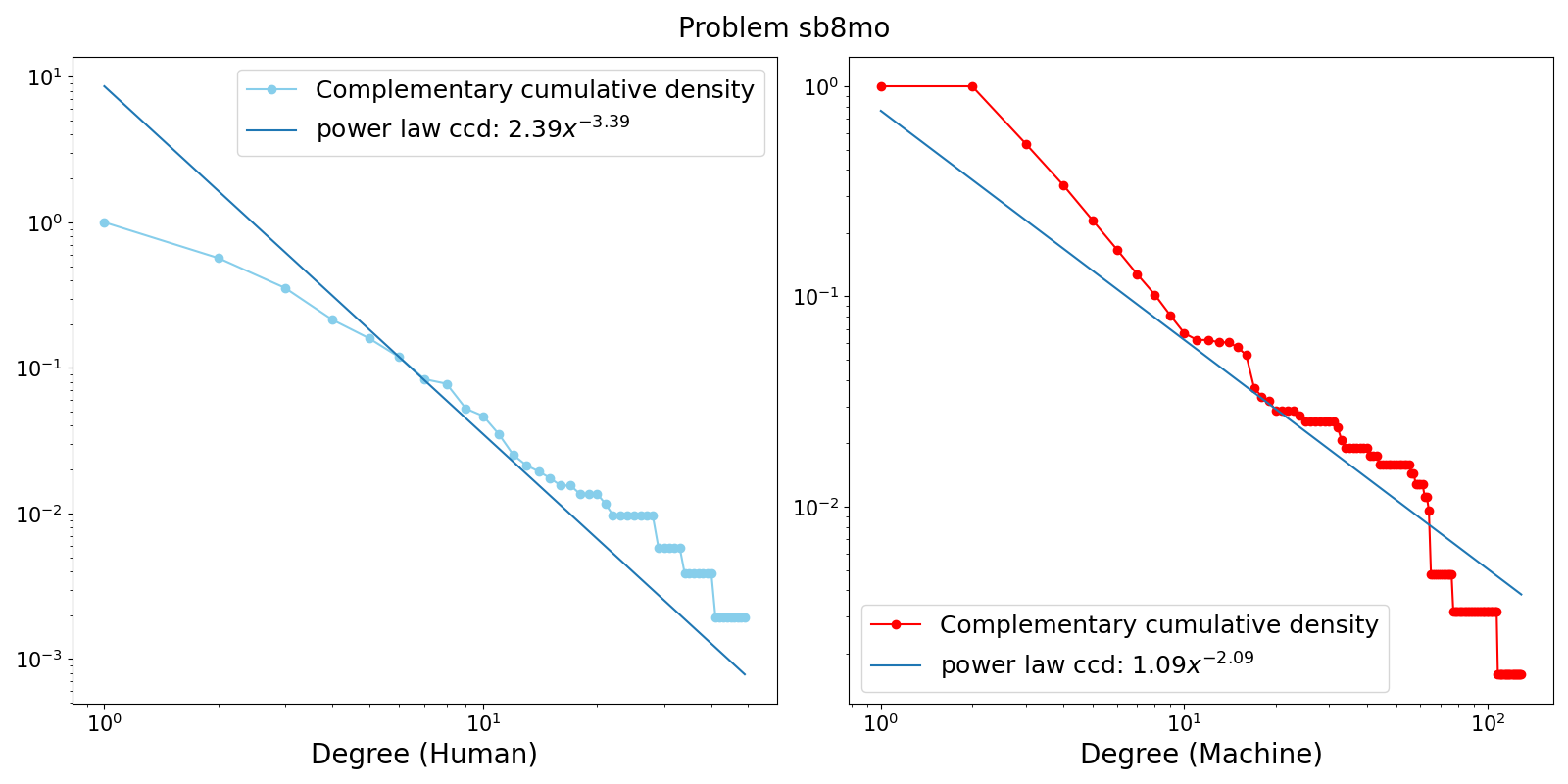}
    \includegraphics[width=0.32\linewidth]{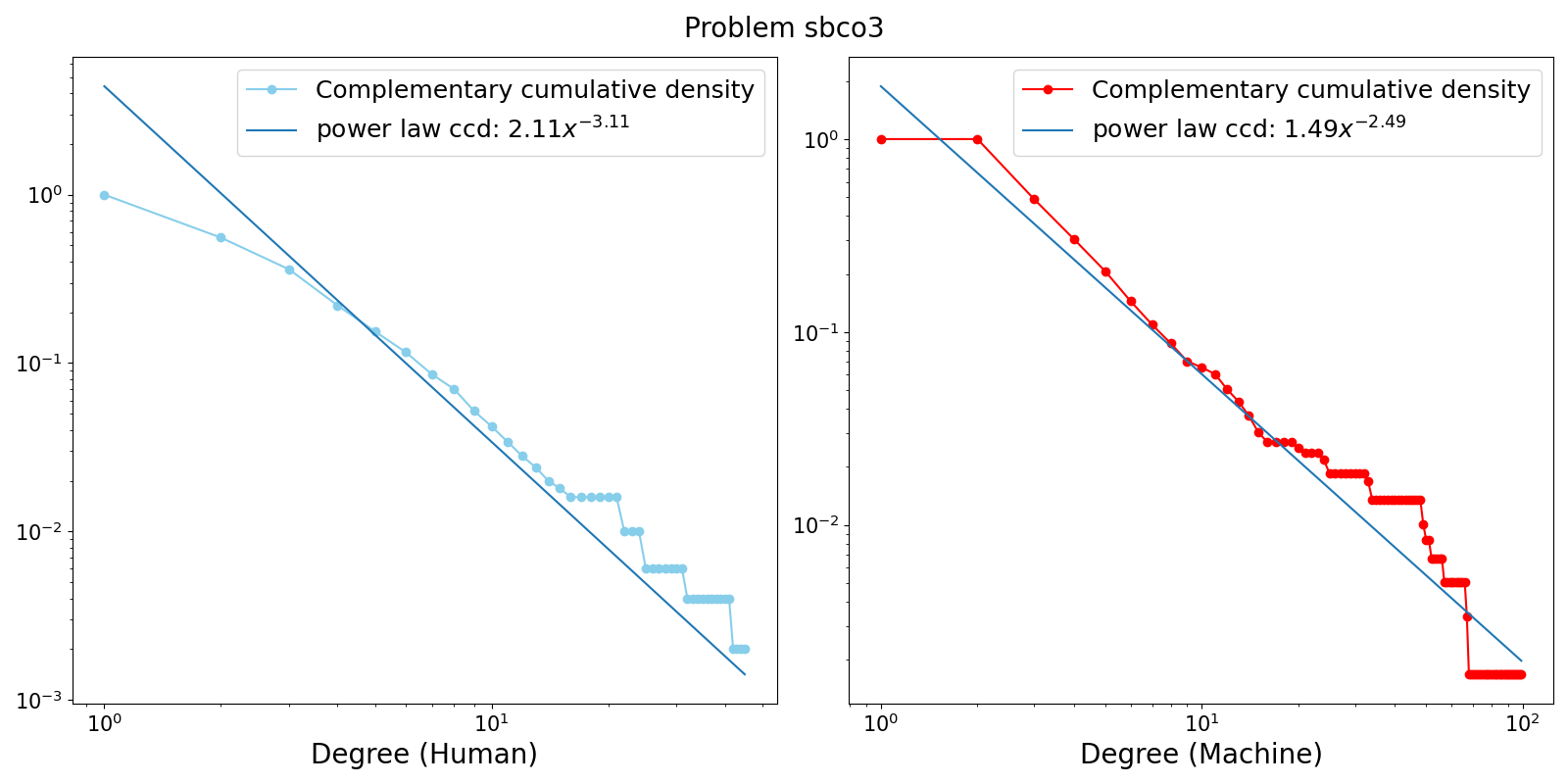}
    \includegraphics[width=0.32\linewidth]{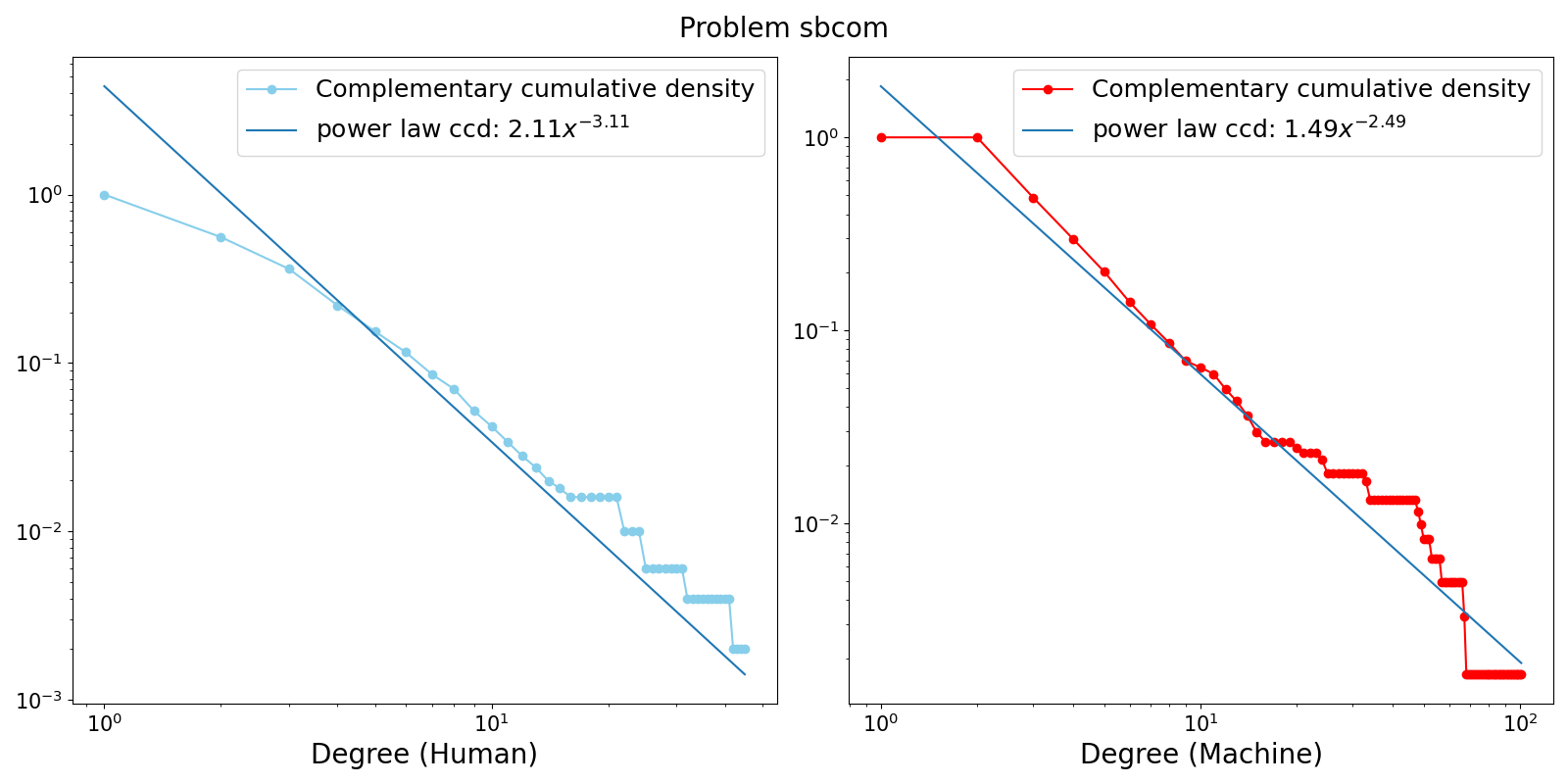}
    
  \caption{Complementary cumulative distribution of the empirical in-degree
    distribution, along with that of the fitted scale-free graph for selected
    theorems of \setmm. Both axes are logarithmic.}
  \label{fig:power_law_fit_app}
  \end{center}
\end{figure*}

Figure~\ref{fig:power_law_fit_app} supports observation~\ref{obs-scalefree} in
Sect.~\ref{sec-comparing-mini} with additional visualizations of the in-degree
distribution of \PDNETs. Here they are considered for grammars obtained for
single theorems of \setmm. They contain the theorem's proof represented as a
production, supplemented by productions representing the proofs of the
theorems that are directly or indirectly referenced in it.

\section{Compressing Human Proofs with New Lemmas}
\label{app:human_compress}

\name{Metamath Proof Explorer} is the main database of \MM. It is also known
by its file name as \name{set.mm}. Aside from comments, declarations and
axioms it contains 43,920 theorems with verified proofs (as of January 2025).
As a KB, its proof grammar thus has 43,920 productions. \setmm can be obtained
as a single file in which the theorems are ordered ``bottom-up'', i.e., a
theorem is proven from axioms and theorems that appear earlier, as assumed
here for sequence ordering of grammars. We now focus on $\SETCORE$, described
in Sect.~\ref{sec-properties-kbs}, which comprises the initial 60\% of \setmm.
It starts with propositional and predicate calculus, moves on to set theory
and then develops 18 mathematical topics.

As described in Sect.~\ref{sec-humancompress}, human-curated grammars can be
further reduced via our TreeRePair-based compression pipeline. We perform this
compression separately for each topic in \SETCORE and provide original, as
well as reduced grammar sizes in Table~\ref{tab:human_compress}. The
compression for topic \name{Elementary Geometry} did not terminate in 7 days
and hence is left out of the total count. For the other topics, running time
varied between 5 minutes to 5 days.

\begin{table}[htb]
  \caption{For each topic $\mathcal{T}$ in \SETCORE, we extract the
    corresponding grammar $G_{\mathcal{T}}$ and compress it via TreeRePair to
    obtain grammar $G_{\mathcal{T}}^{\TRP}$. We show how the number of
    productions and the grammar size change in the compression, and, for the
    grammar size also the percentage of the reduction.}
  \label{tab:human_compress}
  \begin{center}
    \setlength{\tabcolsep}{3pt}
    \begin{tabular}{lrrrrr }\toprule
    Topic $\mathcal{T}$ & $\N(G_{\mathcal{T}})$ & $\N(G_{\mathcal{T}}^{\TRP})$ & $|G_{\mathcal{T}}|$ & $|G_{\mathcal{T}}^{\TRP}|$ &  Reduction \\
    \midrule
    Propositional calculus & 1,743 & 1,843 & 5,760 & 5,191 & 10\% \\
    Predicate calculus & 904 & 1,050 & 4,357 & 3,942 & 10\% \\
    Zermelo-Fraenkel set theory & 2,436 & 2,964 & 16,343 & 14,887 & 9\% \\
    Axiom of replacement & 4,831 & 5,753 & 121,521 & 115,476 & 5\% \\
    Axiom of choice & 240 & 785 & 21,693 & 16,551 & 24\% \\
    Tarski-Grothendieck set theory & 147 & 356 & 4,986 & 3,469 & 30\% \\
    Real and complex numbers & 5,087 & 5,986 & 174,933 & 163,507 & 7\% \\
    Elementary number theory & 852 & 1,716 & 96,692 & 88,497 & 8\% \\
    Basic structures & 452 & 804 & 10,456 & 8,925 & 15\%\\
    Basic category theory & 543 & 1230 & 57,242 & 51,440 & 10\% \\
    Basic order theory & 280 & 516 & 7,278 & 6,112 & 16\% \\
    Basic algebraic structures & 2,401 & 3,318 & 169,352 & 163,547 & 4\%\\
    Basic linear algebra & 1,018 & 1,843 & 88,757 & 79,573 & 10\% \\
    Basic topology & 2,380 & 3,296 & 171,054 & 162,217 & 5\% \\
    Basic real and complex analysis & 581 & 1,451 & 193,877 & 182,005 & 6\%\\
    Basic real and complex functions & 1,501 & 2,378 & 499,438 & 459,803 & 8\% \\
    Elementary geometry & [514] & -- & [139,192] & -- & -- \\
    Graph theory & 1,324 & 2,193 & 41,904 & 37,701 & 10\% \\
    \midrule
    Total & 26720 & 37482 & 1,685,643 & 1,562,843 & 7\% \\    
    \bottomrule
  \end{tabular}
  \end{center}
\end{table}

In the following, we list for each topic of \SETCORE the top 5 proposed lemmas
according to their save-value (we do not show lemmas with a save-value less
than 10). Most of these lemmas are simple combinations of two existing
theorems, however, there are more complex ones as well. We also list some of
these. For each lemma, we provide its save-value, the size of its proof term
and the proven formula, which is the grammar-MGT of the proof term with
respect to the theorem names and statements from \SETCORE taken as
presupposition base. Lemmas that are subsumed by existing theorems are now shown.

\subsection{Propositional calculus}
\begin{Verbatim}[fontsize=\small,breaklines=true,breaksymbolleft=\,,breaksymbolindentnchars=6]
------------
lemma275(A) ->
d(d(d(d(imim1, imim1), d(imim1, imim1)),
    d(d(imim1, 'ax-1'), d(d(imim1, imim1), merlem11))),
  d(d(imim1, imim1), A)).

Save-value: 17, proof term size: 22

$e |- ( ( ( A -> B ) -> ( C -> B ) ) -> ( D -> E ) ) $.
$p |- ( D -> ( ( C -> A ) -> E ) ) $.
------------
lemma33(A) -> d(merco1, A).

Save-value: 16, proof term size: 2

$e |- ( ( ( ( A -> B ) -> ( C -> F. ) ) -> D ) -> E ) $.
$p |- ( ( E -> A ) -> ( C -> A ) ) $.
------------
lemma123 -> d(tbwlem1, imim2i('tbw-ax4')).

Save-value: 12, proof term size: 3

$p |- ( A -> ( ( A -> F. ) -> B ) ) $.
------------
\end{Verbatim}

\subsection{Predicate calculus}

\begin{Verbatim}[fontsize=\small,breaklines=true,breaksymbolleft=\,,breaksymbolindentnchars=6]
------------
lemma351(A) -> '3bitr4g'(notbid(A), 'df-ex', 'df-ex').

Save-value: 14, proof term size: 4

$e |- ( A -> ( A. B -. C <-> A. D -. E ) ) $.
$p |- ( A -> ( E. B C <-> E. D E ) ) $.
------------
\end{Verbatim}

\subsection{Zermelo-Fraenkel set theory}

\begin{Verbatim}[fontsize=\small,breaklines=true,breaksymbolleft=\,,breaksymbolindentnchars=6]
------------
lemma315(A) -> nfan(nfv, A).

Save-value: 24, proof term size: 2

$e |- F/ A B $.
$p |- F/ A ( C /\ B ) $.
------------
lemma347(A) -> anbi12d(eleq1, A).

Save-value: 20, proof term size: 2

$e |- ( A = B -> ( C <-> D ) ) $.
$p |- ( A = B -> ( ( A e. E /\ C ) <-> ( B e. E /\ D ) ) ) $.
------------
lemma1 -> elv(eliin).

Save-value: 18, proof term size: 1

$p |- ( A e. |^|_ B e. C D <-> A. B e. C A e. D ) $.
------------
lemma317(A) -> syl(eqcomd(iftrue), A).

Save-value: 16, proof term size: 3

$e |- ( A = if ( B , A , C ) -> D ) $.
$p |- ( B -> D ) $.
------------
\end{Verbatim}
\begin{Verbatim}[fontsize=\fontsize{8pt}{9pt}\selectfont]
lemma606(A) ->
impbii(orrd(syl6ibr(anc2li(A), eqss)), jaoi(mpbiri('0ss', sseq1), eqimss)).
\end{Verbatim}        
\begin{Verbatim}[fontsize=\small,breaklines=true,breaksymbolleft=\,,breaksymbolindentnchars=6]
Save-value: 15, proof term size: 10

$e |- ( A C_ B -> ( -. A = (/) -> B C_ A ) ) $.
$p |- ( A C_ B <-> ( A = (/) \/ A = B ) ) $.
------------
lemma532(A) -> orbi12d(eqeq2, eqeq1d(ineq2d(A))).

Save-value: 11, proof term size: 4

$e |- ( A = B -> C = D ) $.
$p |- ( A = B -> ( ( E = A \/ ( F i^i C ) = G ) <-> ( E = B \/ ( F i^i D ) = G ) ) ) $.
------------
\end{Verbatim}

\subsection{Axiom of replacement}

\begin{Verbatim}[fontsize=\small,breaklines=true,breaksymbolleft=\,,breaksymbolindentnchars=6]
------------
lemma1 -> a1i(omelon).

Save-value: 123, proof term size: 1

$p |- ( A -> _om e. On ) $.
------------
lemma2 -> adantl(eldifi).

Save-value: 81, proof term size: 1

$p |- ( ( A /\ B e. ( C \ D ) ) -> B e. C ) $.
------------
lemma4 -> fveq2d(suceq).

Save-value: 64, proof term size: 1

$p |- ( A = B -> ( C ` suc A ) = ( C ` suc B ) ) $.
------------
lemma5 -> a1i('0ex').

Save-value: 60, proof term size: 1

$p |- ( A -> (/) e. _V ) $.
------------
lemma6 -> oveq2d(oveq2).

Save-value: 59, proof term size: 1

$p |- ( A = B -> ( C D ( E F A ) ) = ( C D ( E F B ) ) ) $.
------------
lemma351 -> '3adant2'(f1oeq1d(uneq2d(preq1d(opeq2d(f1ocnvfv2))))).

Save-value: 25, proof term size: 5

$p |- ( ( A : B -1-1-onto-> C /\ D /\ E e. C ) -> ( ( F u. { <. G , ( A ` ( `' A ` E ) ) >. , H } ) : I -1-1-onto-> J <-> ( F u. { <. G , E >. , H } ) : I -1-1-onto-> J ) ) $.
------------
lemma131 -> fveq1d(fveq2d(fdmd(ad2antll(elmapi)))).

Save-value: 20, proof term size: 4

$p |- ( ( A /\ ( B /\ C e. ( D ^m E ) ) ) -> ( ( F ` dom C ) ` G ) = ( ( F ` E ) ` G ) ) $.
------------
lemma441 -> fveq2d(mpteq2dv(ifbid(sseq2d(fveq2)))).

Save-value: 12, proof term size: 4

$p |- ( A = B -> ( C ` ( D e. E |-> if ( F C_ ( G ` A ) , H , I ) ) ) = ( C ` ( D e. E |-> if ( F C_ ( G ` B ) , H , I ) ) ) ) $.
------------
\end{Verbatim}

\subsection{Axiom of choice}

\begin{Verbatim}[fontsize=\small,breaklines=true,breaksymbolleft=\,,breaksymbolindentnchars=6]
------------
lemma905(A) -> ad2antrr(syl(A, necon2ai(mtbii(sdom0, breq2)))).

Save-value: 224, proof term size: 6

$e |- ( A -> B ~< C ) $.
$p |- ( ( ( A /\ D ) /\ E ) -> (/) =/= C ) $.
------------
lemma871 -> iunex(omex, ovex).

Save-value: 146, proof term size: 2

$p |- U_ A e. _om ( B C D ) e. _V $.
------------
lemma922 ->
rabbidv(anbi12d(eqeq2d(syl(dmeq, suceq)), eqeq12d(reseq2d(dmeq), id))).

Save-value: 135, proof term size: 9

$p |- ( A = B -> { C e. D | ( E = suc dom A /\ ( F |` dom A ) = A ) } = { C e. D | ( E = suc dom B /\ ( F |` dom B ) = B ) } ) $.
------------
lemma521 ->
anbi12d(eleq12d(id, dmeqd(fveq2)), imbi12d(eleq2, sseq2d(fveq2))).

Save-value: 92, proof term size: 8

$p |- ( A = B -> ( ( A e. dom ( C ` A ) /\ ( D e. A -> E C_ ( F ` A ) ) ) <-> ( B e. dom ( C ` B ) /\ ( D e. B -> E C_ ( F ` B ) ) ) ) ) $.
------------
lemma946(A) ->
eqeltrid('df-ov',
         ffvelcdmd(ad2antlr(f1f), eleqtrrdi(sylibr(simpr, opabidw), A))).

Save-value: 90, proof term size: 9

$e |- A = { <. B , C >. | D } $.
$p |- ( ( ( E /\ F : A -1-1-> G ) /\ D ) -> ( B F C ) e. G ) $.
------------
\end{Verbatim}
\begin{Verbatim}[fontsize=\fontsize{8pt}{9pt}\selectfont]
lemma972(A) ->
adantr(jctir(syl(mpan2(peano1, ffvelcdm),
                 simpld(sylib(sseli(A),
                              elab(fvex,
                                   anbi12d(eleq2d(dmeq), eleq1d(dmeq)))))),
             'pm2.21i'(noel))).
\end{Verbatim}        
\begin{Verbatim}[fontsize=\small,breaklines=true,breaksymbolleft=\,,breaksymbolindentnchars=6]
Save-value: 27, proof term size: 18

$e |- A C_ { B | ( C e. dom B /\ dom B e. D ) } $.
$p |- ( ( E : _om --> A /\ F ) -> ( C e. dom ( E ` (/) ) /\ ( G e. (/) -> H ) ) ) $.
------------
\end{Verbatim}
\begin{Verbatim}[fontsize=\fontsize{8pt}{9pt}\selectfont]
lemma964(A) ->
eleq2d(simpld(simplbda(bitrdi(eleq2d(A),
                              elrab(anbi12d(eqeq1d(dmeq), eqeq1d(reseq1))))))).

Save-value: 13, proof term size: 11

$e |- ( A -> B = { C e. D | ( dom C = E /\ ( C |` F ) = G ) } ) $.
$p |- ( ( A /\ H e. B ) -> ( I e. dom H <-> I e. E ) ) $.
------------
\end{Verbatim}

\subsection{Tarski-Grothendieck set theory}

\begin{Verbatim}[fontsize=\fontsize{8pt}{9pt}\selectfont]
------------
lemma502(A) ->
frsucmpt2(A,
          uneq12d(uneq12d(id, unieq),
                  eqtrid(cbviunv(uneq12d(preq12d(pweq, unieq),
                                         rneqd(eqtrid(cbvmptv(preq2),
                                                      mpteq2dv(preq1))))),
                         iuneq12d(id, uneq2d(rneqd(mpteq1))))),
          uneq12d(uneq12d(id, unieq), iuneq12d(id, uneq2d(rneqd(mpteq1))))).
\end{Verbatim}
\begin{Verbatim}[fontsize=\small,breaklines=true,breaksymbolleft=\,,breaksymbolindentnchars=6]
Save-value: 29, proof term size: 31

$e |- A = ( rec ( ( B e. _V |-> ( ( B u. U. B ) u. U_ C e. B ( { ~P C , U. C } u. ran ( D e. B |-> { C , D } ) ) ) ) , E ) |` _om ) $.
$p |- ( ( F e. _om /\ ( ( ( A ` F ) u. U. ( A ` F ) ) u. U_ G e. ( A ` F ) ( { ~P G , U. G } u. ran ( H e. ( A ` F ) |-> { G , H } ) ) ) e. I ) -> ( A ` suc F ) = ( ( ( A ` F ) u. U. ( A ` F ) ) u. U_ G e. ( A ` F ) ( { ~P G , U. G } u. ran ( H e. ( A ` F ) |-> { G , H } ) ) ) ) $.
------------
\end{Verbatim}
\begin{Verbatim}[fontsize=\fontsize{7pt}{8pt}\selectfont]
lemma781 ->
sylancl(syl2anc(biimpar(necon3bid(cardeq0)),
                adantr(ralrimiva(sylancr(canth2(vpwex),
                                         syl2anc(sylc(simpl,
                                                      syldan(syl3anc(simpl,
                                                                     adantl(oneli(cardon)),
                                                                     adantl(cardsdomelir),
                                                                     tskord),
                                                             syldan(tskpw,
                                                                    tskpwss)),
                                                      ssdomg),
                                                 adantr(ensymd(cardidg)),
                                                 domentr),
                                         sdomdomtr))),
                sylan2(d(inawinalem, cardon), mp3an2(cardon, winainflem))),
        cardidm,
        cardaleph).
\end{Verbatim}        
\begin{Verbatim}[fontsize=\small,breaklines=true,breaksymbolleft=\,,breaksymbolindentnchars=6]
Save-value: 26, proof term size: 39

$p |- ( ( A e. Tarski /\ A =/= (/) ) -> ( card ` A ) = ( aleph ` |^| { B e. On | ( card ` A ) C_ ( aleph ` B ) } ) ) $.
------------
lemma641(A) ->
adantr(eqeltrid(A,
                sylanbrc(sylanblrc(sylancl(grutr, tron, trin),
                                   mp2(inss2, epweon, wess),
                                   'df-ord'),
                         inex1g,
                         elon2))).

Save-value: 20, proof term size: 15

$e |- A = ( B i^i On ) $.
$p |- ( ( B e. Univ /\ C ) -> A e. On ) $.
------------
\end{Verbatim}
\begin{Verbatim}[fontsize=\tiny]
lemma942(A, B) ->
syl2anc(adantr(eqeltrid(A,
                        sylanbrc(sylanblrc(sylancl(grutr, tron, trin),
                                           mp2(inss2, epweon, wess),
                                           'df-ord'),
                                 inex1g,
                                 elon2))),
        mpbid(syl3anc(impcom(sylbi(n0,
                                   exlimiv(expcom(ne0d(eleqtrrdi(sylanblrc(mp3an3('0ss',
                                                                                  gruss),
                                                                           '0elon',
                                                                           elin),
                                                                 A)))))),
                      adantr(eqeltrid(A,
                                      sylanbrc(sylanblrc(sylancl(grutr,
                                                                 tron,
                                                                 trin),
                                                         mp2(inss2,
                                                             epweon,
                                                             wess),
                                                         'df-ord'),
                                               inex1g,
                                               elon2))),
                      adantr(sylc(eqeltrid(A,
                                           sylanbrc(sylanblrc(sylancl(grutr,
                                                                      tron,
                                                                      trin),
                                                              mp2(inss2,
                                                                  epweon,
                                                                  wess),
                                                              'df-ord'),
                                                    inex1g,
                                                    elon2)),
                                  ralrimiva(sylan2(sseli(eqsstri(A, inss1)),
                                                   sylancr(canth2(vpwex),
                                                           sylancr(ensymi(cardid(pwex(vpwex))),
                                                                   sylc(adantr(eqeltrid(A,
                                                                                        sylanbrc(sylanblrc(sylancl(grutr,
                                                                                                                   tron,
                                                                                                                   trin),
                                                                                                           mp2(inss2,
                                                                                                               epweon,
                                                                                                               wess),
                                                                                                           'df-ord'),
                                                                                                 inex1g,
                                                                                                 elon2))),
                                                                        syldan(syldan(grupw,
                                                                                      grupw),
                                                                               sylc(adantr(eqeltrid(A,
                                                                                                    sylanbrc(sylanblrc(sylancl(grutr,
                                                                                                                               tron,
                                                                                                                               trin),
                                                                                                                       mp2(inss2,
                                                                                                                           epweon,
                                                                                                                           wess),
                                                                                                                       'df-ord'),
                                                                                                             inex1g,
                                                                                                             elon2))),
                                                                                    sylancl(mpanr2(d(endom,
                                                                                                     cardid(pwex(vpwex))),
                                                                                                   mp3an2(cardon,
                                                                                                          grudomon)),
                                                                                            cardon,
                                                                                            eleqtrrdi(biimpri(elin),
                                                                                                      A)),
                                                                                    onelss)),
                                                                        ssdomg),
                                                                   endomtr),
                                                           sdomdomtr))),
                                  inawinalem)),
                      winainflem),
              B),
        cflm).
\end{Verbatim}        
\begin{Verbatim}[fontsize=\small,breaklines=true,breaksymbolleft=\,,breaksymbolindentnchars=6]
Save-value: 20, proof term size: 140

$e |- A = ( B i^i On ) $.
$e |- ( ( B e. Univ /\ B =/= (/) ) -> ( _om C_ A <-> Lim A ) ) $.
$p |- ( ( B e. Univ /\ B =/= (/) ) -> ( cf ` A ) = |^| { C | E. D ( C = ( card ` D ) /\ ( D C_ A /\ A = U. D ) ) } ) $.
------------
\end{Verbatim}

\subsection{Real and complex numbers}

\begin{Verbatim}[fontsize=\small,breaklines=true,breaksymbolleft=\,,breaksymbolindentnchars=6]
lemma1 -> recnd(abscl).

Save-value: 385, proof term size: 1

$p |- ( A e. CC -> ( abs ` A ) e. CC ) $.
------------
lemma2 -> a1i(neg1ne0).

Save-value: 198, proof term size: 1

$p |- ( A -> -u 1 =/= 0 ) $.
------------
lemma3 -> a1i(neg1rr).

Save-value: 161, proof term size: 1

$p |- ( A -> -u 1 e. RR ) $.
------------
lemma4 -> rexbidv(eqeq1).

Save-value: 111, proof term size: 1

$p |- ( A = B -> ( E. C e. D A = E <-> E. C e. D B = E ) ) $.
------------
lemma6 -> eleq1d(csbeq1a).

Save-value: 100, proof term size: 1

$p |- ( A = B -> ( C e. D <-> [_ B / A ]_ C e. D ) ) $.
------------
lemma43 -> cbvrexvw(rexbidv(eqeq2d(oveq1))).

Save-value: 47, proof term size: 3

$p |- ( E. A e. B E. C e. D E = ( A F G ) <-> E. H e. B E. C e. D E = ( H F G ) ) $.
------------
lemma712 -> anbi1d(bicomd(ad2antrr(xlt0neg2))).

Save-value: 12, proof term size: 3

$p |- ( ( ( A e. RR* /\ B ) /\ C ) -> ( ( -e A < 0 /\ D ) <-> ( 0 < A /\ D ) ) ) $.
------------
\end{Verbatim}

\subsection{Elementary number theory}

\begin{Verbatim}[fontsize=\small,breaklines=true,breaksymbolleft=\,,breaksymbolindentnchars=6]
------------
lemma1 -> a1i('2nn').

Save-value: 377, proof term size: 1

$p |- ( A -> 2 e. NN ) $.
------------
lemma2 -> eqeq1d(oveq1).

Save-value: 119, proof term size: 1

$p |- ( A = B -> ( ( A C D ) = E <-> ( B C D ) = E ) ) $.
------------
lemma3 -> a1i('2ne0').

Save-value: 114, proof term size: 1

$p |- ( A -> 2 =/= 0 ) $.
------------
lemma4 -> a1i('1nn0').

Save-value: 100, proof term size: 1

$p |- ( A -> 1 e. NN0 ) $.
------------
lemma5 -> zred(simpl).

Save-value: 94, proof term size: 1

$p |- ( ( A e. ZZ /\ B ) -> A e. RR ) $.
------------
lemma311 -> mpteq2dv(fveq2d(oveq2d(oveq2d(oveq1d(oveq1))))).

Save-value: 21, proof term size: 5

$p |- ( A = B -> ( C e. D |-> ( E ` ( F G ( H I ( ( A J K ) L M ) ) ) ) ) = ( C e. D |-> ( E ` ( F G ( H I ( ( B J K ) L M ) ) ) ) ) ) $.
------------
\end{Verbatim}

\subsection{Basic structures}

\begin{Verbatim}[fontsize=\small,breaklines=true,breaksymbolleft=\,,breaksymbolindentnchars=6]
------------
lemma759(A, B) -> syl(opelxpd(A, B), ffvelcdmi(d(f1of, xpsff1o2(eqid)))).

Save-value: 23, proof term size: 8

$e |- ( A -> B e. C ) $.
$e |- ( A -> D e. E ) $.
$p |- ( A -> ( ( F e. C , G e. E |-> { <. (/) , F >. , <. 1o , G >. } ) ` <. B , D >. ) e. ran ( F e. C , G e. E |-> { <. (/) , F >. , <. 1o , G >. } ) ) $.
------------
\end{Verbatim}
\begin{Verbatim}[fontsize=\fontsize{8pt}{9pt}\selectfont]
lemma783(A, B) ->
sylib(ralrimivw(ralrimivw(mpbird(fmpttd(a1i(mpbir(sstri(ovssunirn,
                                                        mp2b(sstri(strfvss(A),
                                                                   mp2b(fvssunirn,
                                                                        rnss,
                                                                        uniss)),
                                                             rnss,
                                                             uniss)),
                                                  elpw(ovex)))),
                                 B))),
      fmpo(eqid)).
\end{Verbatim}        
\begin{Verbatim}[fontsize=\small,breaklines=true,breaksymbolleft=\,,breaksymbolindentnchars=6]
Save-value: 22, proof term size: 23

$e |- A = Slot B $.
$e |- ( C -> ( D e. E <-> ( F e. G |-> ( H ( A ` ( I ` J ) ) K ) ) : G --> ~P U. ran U. ran U. ran I ) ) $.
$p |- ( C -> ( L e. M , N e. O |-> D ) : ( M X. O ) --> E ) $.
------------
lemma491 ->
imbi2d(albidv('2ralbidv'(imbi1d('3anbi1d'(orbi12d(breq2, breq2)))))).

Save-value: 21, proof term size: 7

$p |- ( A = B -> ( ( C -> A. D A. E e. F A. G e. H ( ( ( I J A \/ K L A ) /\ M /\ N ) -> O ) ) <-> ( C -> A. D A. E e. F A. G e. H ( ( ( I J B \/ K L B ) /\ M /\ N ) -> O ) ) ) ) $.
------------
lemma820(A, B) -> sseqtrdi(syl(eqsstrd(A, B), dmss), d(dmxp, vn0)).

Save-value: 16, proof term size: 8

$e |- ( A -> B = C ) $.
$e |- ( A -> C C_ ( D X. _V ) ) $.
$p |- ( A -> dom B C_ D ) $.
------------
lemma693(A, B) -> sylc(A, eqeltrdi(B, fvex), erex).

Save-value: 16, proof term size: 5

$e |- ( A -> B Er C ) $.
$e |- ( A -> C = ( D ` E ) ) $.
$p |- ( A -> B e. _V ) $.
------------
lemma1002(A) ->
'3anbi123d'(orbi12d(breq1d(simplr), breq1d(simpr)),
            sseq12d(simplr, fveq2d(uneq12d(simpr, A))),
            eleq1d(uneq12d(simplr, A))).

Save-value: 12, proof term size: 15

$e |- ( ( ( A /\ B = C ) /\ D = E ) -> F = G ) $.
$p |- ( ( ( A /\ B = C ) /\ D = E ) -> ( ( ( B H I \/ D J K ) /\ B C_ ( L ` ( D u. F ) ) /\ ( B u. F ) e. M ) <-> ( ( C H I \/ E J K ) /\ C C_ ( L ` ( E u. G ) ) /\ ( C u. G ) e. M ) ) ) $.
------------
\end{Verbatim}

\subsection{Basic category theory}

\begin{Verbatim}[fontsize=\small,breaklines=true,breaksymbolleft=\,,breaksymbolindentnchars=6]
 ------------
lemma497(A) -> subcfn(A, eqidd).

Save-value: 106, proof term size: 2

$e |- ( A -> B e. ( Subcat ` C ) ) $.
$p |- ( A -> B Fn ( dom dom B X. dom dom B ) ) $.
------------
lemma504(A) -> ffvelcdmd('3ad2ant1'(A), simp22).

Save-value: 99, proof term size: 3

$e |- ( A -> B : C --> D ) $.
$p |- ( ( A /\ ( E /\ F e. C /\ G ) /\ H ) -> ( B ` F ) e. D ) $.
------------
lemma316(A) -> simpld(syl(A, funcrcl)).

Save-value: 93, proof term size: 3

$e |- ( A -> B e. ( C Func D ) ) $.
$p |- ( A -> C e. Cat ) $.
------------
lemma443(A) -> ffvelcdmd(adantr(A), simpr1).

Save-value: 77, proof term size: 3

$e |- ( A -> B : C --> D ) $.
$p |- ( ( A /\ ( E e. C /\ F /\ G ) ) -> ( B ` E ) e. D ) $.
------------
lemma581 -> cnveqd(oveq12d(fveq2d(simpl), fveq2d(simpr))).

Save-value: 31, proof term size: 5

$p |- ( ( A = B /\ C = D ) -> `' ( ( E ` A ) F ( G ` C ) ) = `' ( ( E ` B ) F ( G ` D ) ) ) $.
------------
lemma367(A) -> simpld(syl(sylib(A, 'df-br'), funcrcl)).

Save-value: 13, proof term size: 5

$e |- ( A -> B ( C Func D ) E ) $.
$p |- ( A -> C e. Cat ) $.
------------
lemma81 -> pwex(uniex(rnex(uniex(rnex(fvex))))).

Save-value: 12, proof term size: 5

$p |- ~P U. ran U. ran ( A ` B ) e. _V $.
------------
\end{Verbatim}

\subsection{Basic order theory}

\begin{Verbatim}[fontsize=\small,breaklines=true,breaksymbolleft=\,,breaksymbolindentnchars=6]
------------
lemma301 -> syl(simpl1, clatl).

Save-value: 26, proof term size: 2

$p |- ( ( ( A e. CLat /\ B /\ C ) /\ D ) -> A e. Lat ) $.
------------
lemma715(A, B) -> syl2anc(simpl1, simpl2, clatlubcl(A, B)).

Save-value: 22, proof term size: 5

$e |- A = ( Base ` B ) $.
$e |- C = ( lub ` B ) $.
$p |- ( ( ( B e. CLat /\ D C_ A /\ E ) /\ F ) -> ( C ` D ) e. A ) $.
------------
lemma716(A, B) -> syl2anc(simpl1, simpl3, clatlubcl(A, B)).

Save-value: 22, proof term size: 5

$e |- A = ( Base ` B ) $.
$e |- C = ( lub ` B ) $.
$p |- ( ( ( B e. CLat /\ D /\ E C_ A ) /\ F ) -> ( C ` E ) e. A ) $.
------------
\end{Verbatim}
\begin{Verbatim}[fontsize=\fontsize{7pt}{8pt}\selectfont]
lemma642(A, B) ->
eqtri(d(A, '0ex'),
      eqtri('df-oprab',
            abf(nex(nex(nex(intnan(mtbir(br0,
                                         breqi(eqtri(d(B, '0ex'),
                                                     eqtri(reseq2i(abf(reu0)),
                                                           res0))))))))))).

\end{Verbatim}        
\begin{Verbatim}[fontsize=\small,breaklines=true,breaksymbolleft=\,,breaksymbolindentnchars=6]
Save-value: 18, proof term size: 22

$e |- ( (/) e. _V -> A = { <. <. B , C >. , D >. | E F G } ) $.
$e |- ( (/) e. _V -> F = ( H |` { I | E! J e. (/) K } ) ) $.
$p |- A = (/) $.
------------
lemma970(A, B) ->
syl12anc(simplll,
         simpr,
         adantr(elpwid(sseldd(ad2antrr(A), simprr))),
         vtocl2(vex,
                vex,
                imbi12d(anbi2d(anbi12d(sseq12, adantl(sseq1))),
                        syl2an(fveq2, fveq2, sseq12)),
                B)).

Save-value: 18, proof term size: 22

$e |- ( A -> B C_ ~P C ) $.
$e |- ( ( A /\ ( D C_ E /\ E C_ C ) ) -> ( F ` D ) C_ ( G ` E ) ) $.
$p |- ( ( ( ( A /\ H ) /\ ( I /\ J e. B ) ) /\ K C_ J ) -> ( F ` K ) C_ ( G ` J ) ) $.
------------
\end{Verbatim}
\begin{Verbatim}[fontsize=\fontsize{8pt}{9pt}\selectfont]
lemma965(A, B) ->
ssexd(a1i(oprabex(fvex, fvex, a1i(moeq), eqid)),
      syl(alrimiv(alrimiv(alrimiv(biimtrid(A,
                                           anim12d1(syl6ibr(ex(B),
                                                            prss(vex, vex)),
                                                    biimpi(eqcom)))))),
          ssoprab2)).
\end{Verbatim}        
\begin{Verbatim}[fontsize=\small,breaklines=true,breaksymbolleft=\,,breaksymbolindentnchars=6]
Save-value: 17, proof term size: 23

$e |- ( A <-> ( B /\ C = D ) ) $.
$e |- ( ( E /\ B ) -> { F , G } C_ ( H ` I ) ) $.
$p |- ( E -> { <. <. F , G >. , D >. | A } e. _V ) $.
------------
\end{Verbatim}

\subsection{Basic algebraic structures}

\begin{Verbatim}[fontsize=\small,breaklines=true,breaksymbolleft=\,,breaksymbolindentnchars=6]
------------
lemma1 -> subgacs(eqid).

Save-value: 284, proof term size: 1

$p |- ( A e. Grp -> ( SubGrp ` A ) e. ( ACS ` ( Base ` A ) ) ) $.
------------
lemma2 -> subgss(eqid).

Save-value: 105, proof term size: 1

$p |- ( A e. ( SubGrp ` B ) -> A C_ ( Base ` B ) ) $.
------------
lemma4 -> eleq1d(oveq2).

Save-value: 75, proof term size: 1

$p |- ( A = B -> ( ( C D A ) e. E <-> ( C D B ) e. E ) ) $.
------------
lemma6 -> adantr(nsgsubg).

Save-value: 70, proof term size: 1

$p |- ( ( A e. ( NrmSGrp ` B ) /\ C ) -> A e. ( SubGrp ` B ) ) $.
------------
lemma5 -> eqeq1d(oveq2).

Save-value: 68, proof term size: 1

$p |- ( A = B -> ( ( C D A ) = E <-> ( C D B ) = E ) ) $.
------------
lemma87 -> raleqbi1dv(notbid(eleq2d(fveq2d(difeq1)))).

Save-value: 14, proof term size: 4

$p |- ( A = B -> ( A. C e. A -. D e. ( E ` ( A \ F ) ) <-> A. C e. B -. D e. ( E ` ( B \ F ) ) ) ) $.
------------
lemma381 -> acsmred(adantr(submacs(eqid))).

Save-value: 12, proof term size: 3

$p |- ( ( A e. Mnd /\ B ) -> ( SubMnd ` A ) e. ( Moore ` ( Base ` A ) ) ) $.
------------
\end{Verbatim}

\subsection{Basic linear algebra}

\begin{Verbatim}[fontsize=\small,breaklines=true,breaksymbolleft=\,,breaksymbolindentnchars=6]
------------
lemma871 -> sselid(ssrab2, simpr).

Save-value: 172, proof term size: 2

$p |- ( ( A /\ B e. { C e. D | E } ) -> B e. D ) $.
------------
lemma748(A) -> a1i(fvexi(A)).

Save-value: 88, proof term size: 2

$e |- A = ( B ` C ) $.
$p |- ( D -> A e. _V ) $.
------------
lemma904 ->
imbi2d(eqeq12d(mpteq2dv(ifbid(eqeq2d(mpteq2dv(ifeq1)))), oveq1)).

Save-value: 12, proof term size: 7

$p |- ( A = B -> ( ( C -> ( D e. E |-> if ( F = ( G e. H |-> if ( I , A , J ) ) , K , L ) ) = ( A M N ) ) <-> ( C -> ( D e. E |-> if ( F = ( G e. H |-> if ( I , B , J ) ) , K , L ) ) = ( B M N ) ) ) ) $.
------------
lemma321 -> cbvralvw(breq1d(mpteq2dv(mpteq2dv(ifeq1d(fveq1))))).

Save-value: 15, proof term size: 5

$p |- ( A. A e. B ( C e. D |-> ( E e. F |-> if ( G , ( A ` H ) , I ) ) ) J K <-> A. L e. B ( C e. D |-> ( E e. F |-> if ( G , ( L ` H ) , I ) ) ) J K ) $.
------------
lemma312 -> eqeq2d(fveq2d(mpoeq3dv(ifeq2d(ifbid(eleq2))))).

Save-value: 12, proof term size: 5

$p |- ( A = B -> ( C = ( D ` ( E e. F , G e. H |-> if ( I , J , if ( K e. A , L , M ) ) ) ) <-> C = ( D ` ( E e. F , G e. H |-> if ( I , J , if ( K e. B , L , M ) ) ) ) ) ) $.
------------
\end{Verbatim}

\subsection{Basic topology}

\begin{Verbatim}[fontsize=\small,breaklines=true,breaksymbolleft=\,,breaksymbolindentnchars=6]
------------
lemma1 -> a1i(iitopon).

Save-value: 180, proof term size: 1

$p |- ( A -> II e. ( TopOn ` ( 0 [,] 1 ) ) ) $.
------------
lemma3 -> cldss(eqid).

Save-value: 76, proof term size: 1

$p |- ( A e. ( Clsd ` B ) -> A C_ U. B ) $.
------------
lemma20 -> adantl(elinel2).

Save-value: 68, proof term size: 1

$p |- ( ( A /\ B e. ( C i^i D ) ) -> B e. D ) $.
------------
lemma5 -> xkotopon(eqid).

Save-value: 62, proof term size: 1

$p |- ( ( A e. Top /\ B e. Top ) -> ( B ^ko A ) e. ( TopOn ` ( A Cn B ) ) ) $.
------------
lemma291 -> rexrd(rehalfcld(rpred(adantr(simpl1r)))).

Save-value: 18, proof term size: 4

$p |- ( ( ( ( ( A /\ B e. RR+ ) /\ C /\ D ) /\ E ) /\ F ) -> ( B / 2 ) e. RR* ) $.
------------
lemma561 -> fveq1d(oveq2d(oveq1d(fveq2d(sneq)))).

Save-value: 15, proof term size: 4

$p |- ( A = B -> ( ( C D ( ( E ` { A } ) F G ) ) ` H ) = ( ( C D ( ( E ` { B } ) F G ) ) ` H ) ) $.
------------
\end{Verbatim}

\subsection{Basic real and complex analysis}

\begin{Verbatim}[fontsize=\small,breaklines=true,breaksymbolleft=\,,breaksymbolindentnchars=6]
------------
lemma979(A) -> syl(A, cncff).

Save-value: 748, proof term size: 2

$e |- ( A -> B e. ( C -cn-> D ) ) $.
$p |- ( A -> B : C --> D ) $.
------------
lemma990(A) -> syl(A, i1ff).

Save-value: 636, proof term size: 2

$e |- ( A -> B e. dom S.1 ) $.
$p |- ( A -> B : RR --> RR ) $.
------------
lemma1 -> necon1ai(ndmioo).

Save-value: 385, proof term size: 1

$p |- ( ( A (,) B ) =/= (/) -> ( A e. RR* /\ B e. RR* ) ) $.
------------
lemma2 -> a1i(reex).

Save-value: 278, proof term size: 1

$p |- ( A -> RR e. _V ) $.
------------
lemma76 -> negnegd(recnd(sselda(a1i(ioossre)))).

Save-value: 29, proof term size: 4

$p |- ( ( A /\ B e. ( C (,) D ) ) -> -u -u B = B ) $.
------------
lemma421 -> mpteq2dva(mulm1d(recnd(ffvelcdmda(i1ff)))).

Save-value: 12, proof term size: 4

$p |- ( A e. dom S.1 -> ( B e. RR |-> ( -u 1 x. ( A ` B ) ) ) = ( B e. RR |-> -u ( A ` B ) ) ) $.
------------
\end{Verbatim}

\subsection{Basic real and complex functions}

\begin{Verbatim}[fontsize=\small,breaklines=true,breaksymbolleft=\,,breaksymbolindentnchars=6]
------------
lemma2 -> adantl(elioore).
Save value: 2209, proof term size: 1

$p |- ( ( A /\ B e. ( C (,) D ) ) -> B e. RR ) $.
------------
lemma1 -> adantl(elfznn).
Save value: 1882, proof term size: 1

$p |- ( ( A /\ B e. ( 1 ... C ) ) -> B e. NN ) $.
------------
lemma11 -> simpld(adantl(eliooord)).
Save value: 1202, proof term size: 2

$p |- ( ( A /\ B e. ( C (,) D ) ) -> C < B ) $.
------------
lemma4 -> a1i('1rp').
Save value: 1048, proof term size: 1

$p |- ( A -> 1 e. RR+ ) $.
------------
lemma5 -> a1i('2re').
Save value: 649, proof term size: 1

$p |- ( A -> 2 e. RR ) $.
------------
lemma601 -> adantrl(biimpa('3adant1'(bicomd(necon3bid(subeq0))))).
Save value: 27, proof term size: 5

$p |- ( ( ( A /\ B e. CC /\ C e. CC ) /\ ( D /\ B =/= C ) ) -> ( B - C ) =/= 0 ) $.
------------
lemma516 -> nnne0d(peano2nnd(imp(ad2antrl(ex(eluznn))))).
Save value: 11, proof term size: 5

$p |- ( ( ( A /\ ( B e. NN /\ C ) ) /\ D e. ( ZZ>= ` B ) ) -> ( D + 1 ) =/= 0 ) $.
------------
\end{Verbatim}

\subsection{Graph theory}

\begin{Verbatim}[fontsize=\small,breaklines=true,breaksymbolleft=\,,breaksymbolindentnchars=6]
------------
lemma814(A) -> eqtr4di(fveq2, A).

Save-value: 43, proof term size: 2

$e |- A = ( B ` C ) $.
$p |- ( D = C -> ( B ` D ) = A ) $.
------------
lemma813(A) -> com23(ex(A)).

Save-value: 38, proof term size: 2

$e |- ( ( A /\ B ) -> ( C -> D ) ) $.
$p |- ( A -> ( C -> ( B -> D ) ) ) $.
------------
lemma2 -> eleq1d(preq2).

Save-value: 34, proof term size: 1

$p |- ( A = B -> ( { C , A } e. D <-> { C , B } e. D ) ) $.
------------
lemma311 ->
rabbidv(notbid(breq2d(fveq1d(fveq2d(opeq2d(reseq2d(imaeq2d(oveq2)))))))).

Save-value: 24, proof term size: 8

$p |- ( A = B -> { C e. D | -. E F ( ( G ` <. H , ( I |` ( J " ( K L A ) ) ) >. ) ` M ) } = { C e. D | -. E F ( ( G ` <. H , ( I |` ( J " ( K L B ) ) ) >. ) ` M ) } ) $.
------------
\end{Verbatim}

\end{document}